\documentclass[12pt]{article}

\input epsf
\def\beq{\begin{eqnarray}}
\def\eeq{\end{eqnarray}}
\def\m{M_*}
\def\msm{M_{\rm SM}}
\def\mpl{M_{\rm Pl}}

\def\d{\delta^{(N)}(\rho)}
\def\lsim{\mathrel{\rlap{\lower3pt\hbox{\hskip0pt$\sim$}}
    \raise1pt\hbox{$<$}}}         %less than or approx. symbol
\def\gsim{\mathrel{\rlap{\lower4pt\hbox{\hskip1pt$\sim$}}
    \raise1pt\hbox{$>$}}}         %greater than or approx. symbol

\begin{document}

\begin{flushright}
NYU-TH/02/02/10 \\
TPI-MINN-02/4\\
UMN-TH-2044/02 \\
\end{flushright}

\vskip 1cm
\begin{center}
{\Large \bf
Diluting Cosmological Constant
In
\\ 
\vskip 0.2cm

~Infinite Volume Extra Dimensions
}
\vskip 1cm
{Gia Dvali$^a$, Gregory Gabadadze$^b$, and M. Shifman$^b$}

\vskip 1cm
{\it $^a$Department of Physics, New York University, New York, NY 10003\\
$^b$Theoretical Physics Institute, University of Minnesota, Minneapolis,
MN 55455}\\
\end{center}

\vspace{0.9cm}
\begin{center}
{\bf Abstract}
\end{center}

We argue that the cosmological constant problem can be solved
in a braneworld model with {\it infinite-volume} extra dimensions,
avoiding  no-go arguments applicable to theories that are
four-dimensional in the infrared. 
Gravity on the brane becomes higher-dimensional
at super-Hubble distances, which
entails that the relation between the acceleration
rate and vacuum energy density flips upside down compared to
the conventional one. The acceleration rate decreases
with increasing  the energy density. The 
experimentally acceptable rate is obtained for 
the energy density larger than (1 TeV)$^4$.
The results are stable under quantum corrections
because supersymmetry is broken only on the brane and stays 
exact in the bulk of infinite volume extra space. 
Consistency of 4D gravity and cosmology 
on the brane requires the quantum gravity scale to be 
around  $10^{-3}$ eV. Testable predictions emerging within 
this approach are: (i) simultaneous 
modifications of gravity at  sub-millimeter and the Hubble scales; 
(ii) Hagedorn-type saturation in TeV energy collisions due to 
the Regge spectrum with the spacing equal to $10^{-3}$ eV.

\vspace{0.1in}

\newpage

\section{Introduction}

The most severe cosmological problem, from the perspective 
of  an effective quantum field
theory,  is undoubtedly the one of the
{\it cosmological constant}. In the nutshell the problem can
be outlined as follows.
The acceleration rate of the universe, $H$, if nonzero,  
has to be
less than or equal to   the present-day value of the Hubble parameter
$H_0$ \cite{WeinbergPRL}\footnote{We use the term {\it acceleration rate} 
in order to distinguish it from the {\it inflation rate} 
in the early Universe.}:
\beq
H\,\lsim H_0\,\sim\,10^{-33}\,{\rm eV}\, .
\label{Hexp}
\eeq
Recent experimental data \cite {cc} appear to confirm a nonzero
value of $H$ that nearly saturates the upper bound in (\ref {Hexp}).

In general relativity $H^2$ determines the scalar curvature
of space and is related to the vacuum energy density
${\cal E}_4$ as
\beq
\mpl^2 \,H^2\, \sim \,{\cal E}_4\,.
\label{4DH}
\eeq
Here $\mpl\sim 10^{18}~{\rm GeV}$ is the Planck mass.
This relation holds as long as ${\cal E}_4\ll \mpl^4$.
A {\it natural} value of
${\cal E}_4$ due to the zero-point energies of known elementary
particles can be estimated to be
$${\cal E}_4\,\sim \,M_{\rm susy}^4\,\gsim\,\left ({\rm TeV}\right
)^4\,,$$
where $M_{\rm susy}$ is the supersymmetry (SUSY) breaking scale.
Substituting this value of ${\cal E}_4$ into (\ref {4DH}), one  finds
$H \,\gsim\, 10^{-3}\,{\rm eV}$, which is grossly inconsistent
with  (\ref {Hexp})\footnote{If ${\cal E}_4$ is negative it gives
rise to anti de Sitter space which is excluded as well.}.

Since the discrepancy between theory and experiment 
manifests itself already at 
enormous distances $\sim H_0^{-1}\simeq 10^{28}$ cm
(i.e., at extremely low energies), it is natural to suspect 
that the problem is not necessarily due to  our lack of knowledge 
of the underlying short-distance physics.
Instead,  this may indicate that, in the standard picture of
the Universe   at cosmic distances, we may be missing an essential part
of the puzzle. The problem is so severe, that it seems reasonable 
to put aside all the other cosmological issues treating them 
as secondary and focus completely on the cosmological constant problem.
This is the approach we take in the present work.
The question   we address is essentially as follows:
``Can the cosmological constant problem originate from our treatment of  
space-time as   four-dimensional (4D) at cosmic distances, 
while in fact it is not?"
As we shall discuss  below the answer {\em can be positive}.
In particular, we will argue that the problem can be avoided 
in a model of ``brane induced gravity'' \cite {DGP} in which 
a graviton propagates  in extra dimensions that have an {\it
infinite volume}.

In order to explain our main result, let us set the terminology first.
In the brane-world models with {\it finite-volume} extra dimensions 
there exists a normalizable four-dimensional zero-mode graviton   
mediating 4D  gravitational interactions 
at distances larger than the size of the extra space.
The observable  4D Planck mass $\mpl$
(the inverse square root of Newton's constant $G_N$)
is then determined by the
$(4+N)$-dimensional Planck mass $\m$, and
the volume of extra space $V_N$ as
\beq
\mpl^2\,= \, \m^{2+N}\, V_N\,.
\label{KK}
\eeq
The same quantity sets the norm (and, thus, the inverse coupling) of a 
zero-mode graviton. 
Therefore, in theories with finite-volume extra dimensions,
one can give  two  equivalent definitions of the 4D  Planck
mass:

(i) As an  inverse square root of   Newton's constant, defining the
strength of  the 4D  gravitational potential;  

(ii) As the norm (or the inverse coupling) of the massless graviton.

In the latter case the value of $V_N$ is constrained by Eq. (\ref {KK}).
In contrast, such a relation  is not possible when $V_N$ is infinite.
Throughout the paper we shall stick to the first definition of $\mpl$,
since this is the only sensible definition in the models \cite{DGP,DG} 
with the infinite-volume  extra dimensions. In these theories the 4D zero-mode
graviton is non-normalizable and, therefore, it is infinitely weakly coupled
because 
\beq
\mpl^2\, \ne \,\m^{2+N}\,V_N\, \to \, \infty \,.
\label{antiKK}
\eeq
In that approach  \cite{DGP} the four-dimensional Plank mass is
induced by loops of particles  localized on the brane. 
The absolute value of  $\mpl$ is
unrelated to $\m$. Instead, it is determined by  ultraviolet properties
of the matter fields on the brane. 
Hence the name, ``{\em brane induced gravity''}.

The fact that the value of $V_N$  is not tied
to $\mpl$ will be crucial  in what follows.
The laws of 4D gravity in this model are obtained only at distances
shorter than the Hubble size.
At larger distances the effects of infinite-volume
extra dimensions take over, and gravity follows  higher-dimensional
laws \cite {DGP}. 

Below we will argue that in the brane induced
gravity model with
$N$ extra spatial dimensions the relation between $H$ and
${\cal E}_4$ (for $N\ne 2$) takes the form
\beq
\m^{2+N}\,H^{2-N}\,\sim\,{\cal E}_4\,.
\label{DH}
\eeq
The key point is that for $N>2$ the acceleration rate $H$ decreases
when 
${\cal E}_4$ is increasing,
\beq
H\,\sim\,\m\,\left (  {\m^4 \over {\cal E}_4  } \right )^{1\over N-2}\,.
\label{HN}
\eeq
This is exactly opposite of the 4D result (\ref {4DH}),
corresponding to  $N=0$, according to which
the rate grows as  ${\cal E}_4$ increases.
As  will be discussed in detail in Sec. 4,
the expression in Eq.~(\ref {HN}) is valid for
$$\m^4 \ll {\cal E}_4 \ll \mpl^4.$$
The expression for $H$ given in Eq.~(\ref {HN}) can yield the acceleration
rate that is consistent with Eq.~(\ref {Hexp}), as we   discuss at length
in the bulk of the paper. Our construction is
an example illustrating that the cosmological constant problem
can be solved, at least in principle, in an effective field theory
approach, provided that gravity becomes soft  above the scale $\m$ --
the property realized in string theory.

The organization of the paper is as follows. 
In Section 2 we summarize arguments why the braneworld  
is a right framework for solving the cosmological 
constant problem. 
In Section 3 we setup the model and describe its
crucial properties. 
In Section 4 we argue that the model gives rise
to the acceleration rate in our 4D world that 
is consistent with the data.
Section 5 is devoted to the issue of UV softening of 
the background solution due to higher-derivative terms. 
In Section 6 we show that 4D laws of gravity are obtained 
on a non-zero tension  brane in infinite-volume  
extra space due to the induced Einstein-Hilbert term.
In Section 7 we discuss no-go theorems
on the cancellation of the cosmological constant.
We argue that a model with infinite-volume extra dimensions  
evades those no-go theorems.
Discussions are given in Section 8.

\section{How do extra dimensions help?}

The questions that an effective field theorist would ask regarding  any
potential solution of the cosmological constant problem are: 

{\bf (i)} Why it is so important to have extra dimensions?

{\bf (ii)} Can the mechanism be understood in the language of a 
4D low-energy effective field theory? 

{\bf (iii)} How does a prospective solution evade  
the usual no-go arguments against  the cancellation of the
vacuum energy contributions coming
from the domain below the supersymmetry breaking scale?

The content of the paper is devoted to the detailed study 
of these issues. However, simple arguments given  
in this section elucidate the key points. 

\vspace{2mm}

The answer to {\bf(i)} is:

\vspace{2mm}

Recall that if there is a vacuum energy density 
${\cal E}_4\, \ge \,{\rm TeV}^4$ 
in a conventional 4D theory
then it unavoidably gives rise to the following scalar 
curvature of space  
\beq
R\,\sim \, {\cal E}_4/\mpl^2 \, \gsim \, (10^{-3}\,{\rm eV})^2~.
\label{Rth}
\eeq
This can be put in the following simple terms:  
The vacuum energy density ${\cal E}_4$
is a source of gravity, and, as such, it has to 
curve the space; the only space in the 4D theory is
the space in which we live. Hence, our space is curved,
as in (\ref {Rth}),  and this is inconsistent with the data. 

However, if there are more than four dimensions the story
could be different. In that case 
${\cal E}_4$ could curve extra dimensions
instead of curving  our 4D space \cite {Wett,RS}. 
This idea is particularly transparent  
for warped geometries \cite {RS}.    
Consider the following $(4+N)$-dimensional interval: 
\beq
ds^2\,=\,A^2(\rho)\,{\bar g}_{\mu\nu}(x)\,dx^\mu dx^\nu\,
-\,B^2(\rho)\,d\rho^2\,-
\,C^2(\rho)\,\rho^2\,d \Omega^2_{N-1}\,,
\label{interval0}
\eeq
where $\mu,\nu =0,1,2,3$, are the indices denoting our 4D world, 
$\rho \equiv \sqrt{\rho_1^2+...+\rho_N^2}$, 
and $\rho_i$'s  denote the extra coordinates.
The curvature 
invariants of the   metric ${\bar g}_{\mu\nu}(x)$ is what we measure 
in our 4D world. There can exist solutions 
to the  $(4+N)$-dimensional Einstein equations 
in the form of (\ref {interval0})
where ${\cal E}_4$ affects strongly the extra space, i.e., 
the functions $A,B$ and $C$, while 
leaving our 4D space almost intact --- the 4D metric
${\bar g}_{\mu\nu}(x)$ remains almost flat.
In this case the energy density ${\cal E}_4$ ``is spent'' totally 
on curving up the extra space rather than on curving 
our 4D space.

\vspace{2mm}

The answers to {\bf (ii)} and {\bf (iii)} are:

\vspace{2mm}

The effective field theory arguments are based on the assumption that
there is a finite number of 4D degrees of freedom below the scale of
the cosmological constant that one wants to neutralize. 
This condition is {\em not}
satisfied in the present model ---  it  is a genuinely high-dimensional 
theory in  the far infrared. Therefore, there is an infinite 
number of degrees of freedom
below any non-zero energy scale. As a result, there is no scale below
which   extra dimensions can be integrated out and the theory   reduced to
a  {\it local} 4D field theory with a finite number of degrees of freedom.
In order to rewrite the model at hand as a theory of a single
4D graviton, at any given scale, we have to integrate out an {\it infinite}
number of lighter modes. As    usually happens  in field theory,   
integrating out the light states we get  {\it non-local interactions}.
Therefore, the resulting model, rewritten as a theory of a 4D graviton,  
will contain generally-covariant but non-local terms. The latter 
dominate the action in the far infrared. 
Of course, in actuality,  the full theory is local --- the apparent 
non-locality is  an artifact of integrating out light modes. It  
tells us that a local $(4+N)$-dimensional theory can be imitated by a
non-local 4D model. The non-local terms modify the effective 4D 
equations and neutralize a large cosmological constant.
As a result, the relation between the acceleration rate
and the vacuum energy density is changed, as was mentioned in 
Sect. 1.

To emphasize the crucial role of the infinite volume,
below  we would like to  summarize no-go arguments 
on why any 4D theory \cite {Weinberg} or a theory that has  
finite-volume extra dimensions cannot solve the 
cosmological constant problem \cite {DG}\footnote{ 
Let us note that these arguments are not applicable
to solutions of anthropic type  (see discussions in Ref. \cite{vilenkin}),
nor to any solution based on the cancellation
of the cosmological term due to some reason hidden in the
ultraviolet (UV) physics. Such a cancellation would
look like fine-tuning from the low-energy
perspective.  We shall not consider this possibility, 
since it would require the detailed understanding of the UV physics
(see, e.g., discussions in Ref. \cite {Dines}). This is 
beyond the scope of the present investigation.}.

Consider a  theory that, below certain scale $M_c$  ($M_c>>H_0$), 
flows to a local {\it four-dimensional} effective field theory 
with a finite number of fields. In particular,   all
Kaluza-Klein (KK) type theories, in which the
 extra space has a finite volume $\sim 1/M_c^N$, are of this type.
Then, below the scale $M_c$, we
can integrate out  heavy physics and write down an effective low-energy
theory, that --- by construction  --- is generally covariant 
and has a finite number of 4D
light degrees of freedom. As is well known,  such theories require 
the existence of a massless spin-2 state, the graviton. 
The graviton, {\em must} be {\it universally} coupled.

This fact has an extremely important consequence: it excludes the
existence
of sectors with unbroken supersymmetry. The reason is simple.
Supersymmetry in the
Standard Model (SM) sector is broken   at the scale 
$M_{{\rm susy}} \sim {\rm TeV}$ or higher.
Since all   states in the theory  couple to gravity with the equal strength,
the lowest possible SUSY breaking scale
in any sector is
\beq
m_{\rm min}\, \sim \, {{\rm TeV}^2 \over M_{\rm Pl}}\, \sim 10^{-3}\,\,  
{\rm eV}\, .
\label{susyscale}
\eeq
This is a bad news because of the following two
reasons:

(i) Eq. (\ref{susyscale}) implies that the loop contribution to the
cosmological
term from each sector is at least as large as 
\beq
\Delta\varepsilon_{\rm min} \sim {{\rm TeV}^4 \over M_{\rm Pl}^2}\,
\Lambda_{\rm cut-off}^2,
\label{susyscal}
\eeq
where $\Lambda_{\rm cut-off}$ is the UV cut-off;

(ii) The scalars with masses $\ll 10^{-3}$ eV 
are needed to cancel the cosmological constant
(other spin fields, due to the requirement of 4D Lorentz invariance,
can only renormalize couplings  in the potential). 
However, because of the above arguments the existence of 
such light scalars is highly unnatural\footnote{
There exist  two possibilities to have naturally light scalars in 
a theory, as pseudo Nambu-Goldstone
bosons (PNGB), or as composite states. However, neither of these
can help to cancel the big cosmological constant. In the case of the 
PNGB, the potential that corresponds to the explicit symmetry
breaking has to be very high in order to cancel the vacuum energy density.
If so, the corresponding PNGB is heavy and becomes 
irrelevant for the low energy dynamics \cite {DV}. 
As to the light composite bosons,  
they should have the mass of the 
order of $\sim H_0$. This would imply that the compositness scale 
is of the same order, and that the size of 
those scalars is generically huge, $\sim H^{-1}_0$. 
However, the latter scalar cannot do the job since they  
are not composites at the observable distances.}.

The following condition can be derived. 
In order to cancel a small change
in the 4D curvature to a given accuracy $H$, by re-adjusting the VEV of a
scalar field, the latter must be as light as $H$. This is obvious:
if the field is much heavier than $H$,
it can be integrated out and cannot
participate in  the low-energy dynamics at this scale. In particular,
its VEV cannot be sensitive to a small change in the expansion rate.
By the same token, the
dynamics of the SM  Higgs VEV is completely
irrelevant for the present-day cosmological expansion.

Thus, to suppress the acceleration rate down to the observable 
value $H_0$, the existence of scalars as light as 
$10^{-33}$ eV is required. (This is a necessary but not
sufficient condition). Now, as we have argued above, in the
theories with finite-volume extra space, all   states 
couple to gravity universally, via $1/\mpl$, and, therefore, 
the lightest scalars have  mass $\sim 10^{-3}$ eV. 
This kills any hope for 
self-adjustment mechanisms in the theories with finite-volume extra
dimensions. It is  
impossible to cancel cosmological term to   accuracy better
than $\sim {\rm TeV}^4$ in the finite-volume models,
unless a severe fine-tuning is invoked. On the contrary, in the 
infinite-volume models the higher-dimensional scalar components of the 
graviton KK modes can and will be arbitrarily light. 
This will lead to a drastic change, as we shall
discuss in the next section.

\section{The model: importance of infinite volume}

 In this section we shall formulate the model, and summarize the main
assumptions. The model is the $N \ge 2 $ version of  
Ref.~\cite {DGP}
(for generalizations to $N \ge 2$ see \cite {DG,Kiritsis},
and Ref.~\cite {DGHS} for a recent summary; for string theory discussions
of this model see Refs. \cite {ZKaku,Kiritsis,Brown}.).

The theory to be considered is a brane-world model embedded in a space with
(asymptotically) flat {\it infinite-volume}
$N$ extra dimensions. All known SM particles are 
localized on the brane and obey the conventional 4D laws 
of gauge interactions up until very high energies,  of the order of 
the GUT scale for instance. The gravitational
sector, on the other hand, is spread over the whole $(4+N)$-dimensional
space.  The low-energy action of the model is written as follows:
\beq
{S}\,=\,
\m^{2+N}\,\int\,d^4x\,d^N\rho\,\sqrt{G}\,{\cal R}\,+\,\int\, d^4x
\sqrt{\bar g}\,\left ({\cal E}_4\,+\,M^2_{\rm ind}\,{\overline R}\,
+\,{\cal L}_{\rm SM}(\Psi, \msm)\right )\,.
\label{actD}
\eeq
Let us discuss  various parts and  parameters of the 
action (\ref{actD}).
${\cal L}_{\rm SM}$ is the Lagrangian for the particle physics including 
all SM fields $\Psi$ 
\footnote{For simplicity of notations we use the convention 
that the whole theory of particles physics
including any (SUSY) GUT or other extensions of the 
Standard Model is denoted by SM.}. 
The parameter $\msm$ denotes the UV cutoff of the SM.  
Up until that scale the SM obeys the conventional 
4D laws. In the present approach $\msm \gg {\rm TeV} \gg \m$.
Moreover, $G_{AB}$  stands for a
$(4+N)$-dimensional graviton $(A,B=0,1,2,...,3+N)$, while $\rho$
are ``perpendicular'' coordinates.
For simplicity  we do not consider brane
fluctuations~\footnote{This limitation could be readily lifted.
Indeed, including the brane fluctuations
would produce a sterile Goldstone boson,
and heavy modes which could manifest themselves only through
generation of an extrinsic curvature term
on the brane. It is easy to see that the latter can be reduced to an
inessential renormalization of constants in
the action (\ref{actD}).}. Thus, the induced metric on the brane is given by
\beq
{\bar g}_{\mu\nu}(x)~\equiv~G_{\mu\nu}(x, \rho_n=0)\,.
\label{ind}
\eeq
Since we discard the brane fluctuations,
the brane can be thought of as
a boundary of the extra space or
an orbifold fixed point
(if so, the Gibbons-Hawking
surface term is implied in the action hereafter).
The brane tension is denoted by  ${\cal E}_4$.

The first term in (\ref{actD}) is the bulk Einstein-Hilbert  action
for $(4 +N)$-dimensional gravity, with  the fundamental scale $\m$.
The expression in  (\ref{actD}) has to be understood
as an effective low-energy Lagrangian valid for graviton momenta
smaller than $\m$. Therefore, in what follows 
we shall imply the presence of an infinite number of 
gauge-invariant high-dimensional
bulk operators suppressed by powers of $\m$. 

The second term in
(\ref{actD}) describes the 4D Einstein-Hilbert (EH) term of the induced
metric. This term plays the crucial role. 
It ensures that at observable distances 
on the brane the laws of 4D gravity are reproduced 
in spite of the fact that there is no localized 
zero-mode graviton.  Its coefficient $M_{\rm ind}$ is another 
parameter of the model. Thus,  the  low-energy action as it 
stands is governed by three parameters 
$\m$, $M_{\rm ind}$ and ${\cal E}_4$.
Let us discuss their natural values separately.

The parameter $M_{\rm ind}$ gets induced  by  the SM
particle loops localized on the brane. 
Such corrections are cut-off by the rigidity scale of the SM, $\msm$,
i.e.,  the scale above which the SM propagators become soft.
In the present approach this scale is taken to be very high, $\gg$ TeV.
In particular, we will take this scale to be comparable with 
the GUT or 4D Planck scale\footnote{We set  the thickness of the
brane $\Delta$ to be determined by the
SM scale, $\Delta \sim \msm^{-1}\,$.
Therefore, all SM particles live 
on the brane and  obey the 4D laws of non-gravitational
interactions,
while gravity can propagate into the $(4+N)$-dimensional
bulk. In particular, the conventional logarithmic gauge coupling unification
is fully preserved, as well as other useful
properties of the SM, such as the seesaw mechanism for neutrinos, etc. 
The fact that all particles other than gravitons and gravitinos
are localized on the brane is certainly welcome also
from the aesthetic standpoint. We assume for convenience 
and clarity that above $\msm$ the SM stops to run logarithmically 
because of the heavy states with masses $\gsim \msm$ which 
together with the SM particles complete the multiplets of  
a certain super-conformal theory.}.
The loops  induce\footnote{$M_{\rm ind}$  
can certainly contain as well the tree-level terms if 
these are present in the original action in the first place.
We will not discriminate between these and induced terms.
$M_{\rm ind}$ will be regarded as a parameter that 
stands in (\ref {actD}).}
the Einstein-Hilbert term in (\ref {actD}),
\beq
M_{\rm ind}^2\,\sqrt{\bar g}\,
{\overline R}({\bar g} )\,,
\label{delta2}
\eeq
where the value of the induced constant $M_{\rm ind}$ is
determined by the relation~\cite {Adler,Zee},
$$M_{\rm ind}^2=i\int d^4x \,x^2\, \langle T(x) \,T(0)
\rangle/96 \,.$$ 
The parameter $M_{\rm ind}$ is proportional
to the scale $\msm$ and to the number of particles in the SM
~\footnote{The scalars and fermions contribute to $M_{\rm ind}$
with positive sign while the gauge fields
with negative sign.}.
Since there are about 60 particles in
the Weinberg-Salam  model, and more are expected in  GUT's,
the value of $M_{\rm ind}$ should be somewhat
larger than $\msm$. In fact, below we define the 4D Planck mass 
as being completely determined by $M_{\rm ind}$:
\beq
\mpl\,\equiv \, M_{\rm ind}\,.
\label{mpl11}
\eeq
Thus, the Planck mass is not a fundamental constant in our approach
but is rather a derived scale.
We see that the SM loop corrections 
are capable of creating the hierarchy $M_{{\rm ind}}/M_*$, even 
if the initial value of $M_{{\rm ind}}/M_*$ was  not that large.  
This hierarchy does not amount to fine tuning, 
since such a separation of scales is stable 
under quantum corrections. 
Indeed, say,  $\m$ gets renormalized by all
possible bulk quantum gravity loops. 
However, there are no SM particles in the bulk 
the only scale in there is $\m$. 
Therefore, any bulk loop  gets cut-off
at the scale $\m$, as it is the fundamental 
gravity scale. While, as we discussed above,
the brane SM loops are cut-off by the higher scale 
$\msm$, and this gives rise to the huge 
value of $M_{\rm ind}$ on the brane. 

Finally, let us discuss the value of the brane tension. To this end, 
we have to specify our assumptions regarding supersymmetry. We assume that
the high-dimensional theory is supersymmetric, and that supersymmetry is
spontaneously broken only on the brane  (such a scenario with a non-BPS 
brane was considered in \cite{mishagia}). 
Non-breaking of supersymmetry in the bulk is only 
possible due to infinite volume of the extra space;
SUSY breaking is not transmitted
from the brane into the bulk \cite{DGPsusy,Wittensusy} since 
the breaking effects are suppressed by an infinite volume factor
~\footnote{In general,
local SUSY  in the bulk does not preclude a negative vacuum energy
density of the order of $\m^{4+N}$.
However, the latter can be forbidden by an unbroken
$R$ symmetry in the bulk. Such a symmetry is often provided by 
string theory.}.
Then, the  bulk cosmological term can be set to zero, without any
fine-tuning. The natural
value of  ${\cal E}_4$ can be as low as TeV$^4$, since 
the brane tension can be protected above this value by 
${\cal N}=1$ supersymmetry (note that 
$M_{\rm ind}$ can only be protected by a conformal invariance 
which we assume is broken at the scale $\msm$). 
All these properties are summarized in Fig. 1.
\begin{figure}
\centerline{\epsfbox{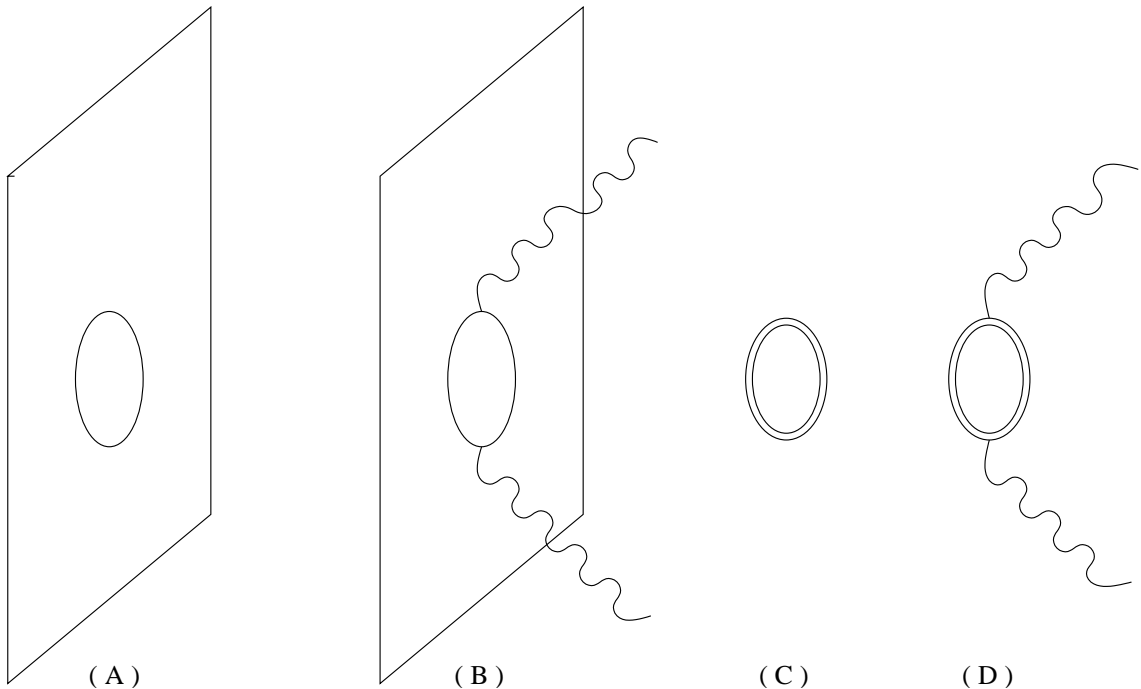}}
\epsfysize=6cm
\vspace{0.5in}
\caption{\small (A) The worldvolume 
vacuum diagram of the SM fields 
that renormalizes the brane tension (i.e., the 4D 
cosmological constant). These contributions are protected  
by ${\cal N}=1$ worldvolume supersymmetry. Therefore, 
they are cutoff by the worldvolume SUSY breaking 
scale $M_{\rm susy}$.
(B) The worldvolume 
two-point diagram that renormalizes (induces) the EH
term on the brane. These contributions {\it are not protected}
by ${\cal N}=1$ supersymmetry. They can only be protected by
conformal invariance which in our model is 
broken at the scale $\msm$ that is close to $\mpl$.
Hence, there can be a hierarchy between (B) and (A).   
(C) The bulk vacuum diagram. Only the bulk particles 
(which do not include the SM particles) are running in this  loop 
(bulk particles are denoted by double lines). 
This diagram is protected by unbroken bulk SUSY. 
Therefore, the cosmological constant in the 
bulk is zero.
(D) The bulk two-point diagram which renormalizes the bulk EH
term. As in (C), only the bulk particles 
are running in the loop. This diagram is cutoff by the bulk 
scale $\m$. Therefore, the natural value of the constant 
in front of the bulk EH term is $\m^{2+N}$; there is huge 
hierarchy between this coefficient and that
of the worldvolume EH term coming from (B).}
\label{fig}
\end{figure} 

Let us now turn to the  gravitational dynamics on the brane.
This dynamics is quite peculiar. Despite the fact that the volume
of extra space is infinite, nevertheless an observer on the brane measures
4D gravitational interaction up to cosmologically large scales \cite{DGP}.
For instance, in the $N\ge 2$ case  
the static potential between two objects on the 
brane scales as
\beq
V(r)\, = \,- {1 \over 16\pi \mpl^2}\,{1\over  r}\,+..\,,
\label{Newton}
\eeq 
for distances in the interval
\beq
\m^{-1}\,\lsim \, r \, \lsim \, r_c \,\sim\,{\mpl / \m^2}\,.
\eeq
However, at distances smaller than $\m^{-1}$ and 
bigger than $r_c$ the potential changes. 
In order for the late-time cosmology to be standard 
we require  that
$r_c ~\sim ~H_0^{-1}\sim 10^{28}~{\rm cm}$.
This unambiguously restricts the value for the quantum gravity scale
around the following value:
$$\m \sim 10^{-3}~ {\rm eV}\,.$$
Small $\m$ means that gravity in the
bulk of extra space is strong. However, the large EH term
on the brane ``shields'' the SM particles on the brane
from strong bulk gravity~\cite {DGKN1}(Small $\m$ was first 
discussed in Ref. \cite {Sundrum} in a different 4D framework,
see also \cite {Moffat})
\footnote{Note that having a small value of $\m$ does not solve 
{\it per se} the cosmological constant problem 
since the SM loops are still cutoff by the high scale.}.  
The case of interest for the cosmological constant problem
is a nonzero tension brane, i.e., $ {\cal E}_4\ne 0$.
For the latter, as we will argue in Section 6, the crossover distance 
is also comparable with the present-day Hubble size.

The ``shielding'' effect plays a crucial role.
We will discuss it in more details here.
The essence of the shielding is as follows. 
Due to the presence of the large induced kinetic term
on the brane, the wave functions of high-dimensional gravitons with large
external momentum are  suppressed on the brane. 
Only the gravitons with extremely low
external momenta (smaller than  $r_c^{-1}$) 
can penetrate to the brane and couple to
the SM particles.
Hence, a brane observer is ``shielded'' from the strong bulk
gravity  and detects a weak
four-dimensional force up until the distances exceeding the crossover scale
$r_c$. The modes participating in gravitational dynamics on the brane at
larger distances are not shielded, and as a result 
gravity becomes high-dimensional beyond the crossover scale.

The effect can be best understood in terms of the four-dimensional
mode expansion. From the four-dimensional perspective a high-dimensional
graviton represents a continuum of four-dimensional states and
can be expanded in these states. Below we shall be interested
only in spin-2 components for which the KK decomposition  can schematically 
be written as follows:
\begin{equation}
  h_{\mu\nu}(x,\rho_n)\, = \,\int d^N m\, \epsilon_{\mu\nu}^{(m)}(x)\,
 \sigma_m (\rho_n)\,,  
\label{kkexp}
\end{equation}
where  $\epsilon_{\mu\nu}^{(m)}(x)$ are four-dimensional spin-2 fields of mass
$m$ and   $\sigma_m(\rho_n)$ are their wave-function profiles
in extra dimensions. The strength of the coupling of an individual mode
to a brane observer is given by the value of the wave-function at the
position of the brane, that is  $\sigma_m(0)$.
 Four-dimensional gravity on the brane is mediated by
exchange of all the above modes. Each of these exchanges mediates 
a Yukawa type gravitational  potential. The result is
\begin{equation}
 V(r)\,\propto \,{1 \over \m^{2 + N}}\,
\int_0^\infty dm\, m^{N-1}\, |\sigma_m(0)|^2
\,{e^{-rm} \over r}\,.
\label{kkpot}
\end{equation}
Four-dimensional gravity on the brane is recovered for $r<<r_c$, due
to the fact that the modes heavier than $1/r_c$ have suppressed wave-functions
and, therefore, the  above integral is effectively cut-off 
at $m \sim 1/r_c$. 
Most easily this can be seen from the propagator analysis.
Gravitational potential (\ref{kkpot}) on the brane is mediated by an
``effective'' 4D-graviton which can be defined as:
\begin{equation}
h_{\mu\nu}(x,0)\, = \,\int d^N m\, \epsilon_{\mu\nu}^{(m)}(x)
\, \sigma_m (0)\,.  
\label{effgraviton}
\end{equation}
The Green's function for this state can be defined in the usual way.
Using (\ref{effgraviton}) and orthogonality of the
$\epsilon_{\mu\nu}^{(m)}(x)$-states we obtain
\begin{equation}
{\cal G}(x-x',0)_{\mu\nu,\gamma\delta}\, = \,
\langle  h_{\mu\nu}(x,0) \,h_{\gamma\delta}(x',0) \rangle =
 \int d^N m |\sigma_m(0)|^2 
 \langle\epsilon_{\mu\nu}^{(m)}(x) 
\epsilon_{\gamma\delta}^{(m)}(x')\rangle \,.
\label{propeffective}
\end{equation}
From now on we shall suppress the tensor structure, which is not
essential for this discussion. Going into the Euclidean
momentum space we get the following expression for the scalar part
of the propagator
\begin{equation}
G(p,0) \, = \,\int dm \, m^{N-1} \,{|\sigma_m (0)|^2 \over m^2 + p^2}\,.
\label{propeffm}
\end{equation}
This is the spectral representation for the
Green's function
\begin{equation}
G(p,0)\,  = \,\int d s \,{\rho(s) \over s + p^2}\,,
\label{KLR}
\end{equation}
with $s \equiv m^2$ and
\begin{equation}
\rho(s) \, = \,{1\over 2}\, s^{{N-2\over 2}} |\sigma_{\sqrt{s}} (0)|^2\,.
\label{rhosigma}
\end{equation}
Therefore, the spectral  representation of the effective graviton
Green's function can be simply understood as the KK mode expansion
(\ref{kkexp}). Thus, the wave-function suppressions for the heavy modes can
be read-off from Eqs. (\ref{propeffm}) and (\ref{KLR}) by using 
the explicit form of the propagator $G(p)$ \cite{DG,DGHS}
\beq
G(p, 0 )\,=\,{1 \over \mpl^2 p^2\,+\,\m^{2+N} D^{-1}(p, 0)}\,,
\label{gee111}
\eeq
where $D^{-1}(p, 0)$ is the inverse Green's function 
of the bulk theory with no brane.
For the purposes of the present discussion it is enough to notice 
that for $N>1$ 
and large momenta $p >> r_c^{-1} = \m^2/\mpl$ 
the above propagator behaves as \cite {DG,DGHS}
\beq
G(p, 0 )\,\simeq\,{1 \over \mpl^2 \,p^2}\,
\label{4dp}
\eeq
which is nothing but the propagator of a 
massless four-dimensional graviton with
the $1/\mpl$ coupling. Substituting (\ref{4dp}) into the left hand side
of (\ref{KLR}), we find  that the function $\rho(s)$ must be
suppressed  for $s>> r_c^{-2}$. If so, the relation (\ref{rhosigma})
implies that the wave-functions of the heavy modes must be 
vanishingly small as well \cite {DG,Wagner}.

For $N=1$ both the propagator \cite{DGP} and the wave-function profiles 
\cite{DGKN1} 
can be evaluated analytically and are given by
\beq
G(p, 0 )\,=\,{1 \over \mpl^2\, p^2\, + \,2\m^3\, p}\,,
\label{N1prop}
\eeq
and
\beq
|\sigma_m(0)|^2\, = \,{4 \over 4 + m^2{\mpl^4 /\m^6}},
\label{N2prof}
\eeq
respectively. This explicitly demonstrates that the modes that are 
heavier than $r^{-1}_c = {\m^3/\mpl}$ 
are effectively shielded (note the difference in
the expression for $r_c$ from $N>1$ case). 
Substituting (\ref{N2prof}) into  (\ref{kkpot}) we derive the
usual Newtonian potential (\ref {Newton}) at distances $r << r_c$.

One can interpret the Green's function
as describing  a metastable state that decays into the bulk states
with the lifetime $\sim r_c$. 
The remarkable fact is that the existence of such  a metastable state
is perfectly  compatible with the  exact 4D general covariance.
At short distances $r \ll r_c$, the theory reproduces all   observed
properties of 4D gravity; it  dramatically differs, however,   
at large distances, $r \gsim r_c $. 
More detailed analysis indicates that all the 
predictions of general relativity,
are reproduced at short distances. The Schwarzschild type behavior of
gravitating bodies at short scales was demonstrated recently in Ref. 
\cite {Gruzinov}.

The no-go arguments discussed in the previous section 
are not applicable to the theories with 
infinite-volume extra dimensions.
The crucial property of this class of theories is 
that despite the unbroken 4D general covariance,
there is no 4D zero-mode graviton.
4D gravity on the brane is mediated by a {\it collective mode}
which  cannot be reduced to any 4D state. 
The fact most important  for us is that the 4D general covariance 
{\it does not} require now all   states to couple universally 
to our ``graviton".  In fact, the
bulk states couple to it infinitely weakly. As a result, there is no
universal agent that could mediate supersymmetry breaking from  
SM to all   existing states. Such a situation is impossible
in the finite-volume theories where 4D gravity is mediated by a normalizable
zero mode, which, by   general covariance, must couple
universally and, hence,  mediates supersymmetry breaking.

Thus, the infinite-volume theories defy both   arguments of Sec. 2:

(i) There can be a sector of the theory with {\it unbroken SUSY};

(ii) Arbitrarily light 4D scalars exist that can be exploited to
neutralize the cosmological term.

Moreover, the effect of the brane cosmological term  is to
excite extra components of the bulk graviton and curve the extra space,
without inducing a large 4D curvature. 
We stress again that this is  impossible in  finite-volume
theories (i.e., the theories in which the size of the extra space
is smaller than the Hubble size $H_0^{-1}$)
because there the extra components of the metric
are always heavier than $H_0$.

Before turning to  the next section let us reiterate the two
main ingredients:

{\it (1) Softness of gravity above $M_*$}. 
If $M_*$ is the scale of quantum gravity in the bulk, we should expect that 
gravitational amplitudes soften for the graviton momenta above $M_*$. 
This is certainly supported by string theory, which at present
is a primary candidate for quantum gravity. However, the softening of 
amplitudes does not come for free. The price to pay is the presence
of an exponential multiplicity of light Regge states in the bulk.
These states have the mass spacing determined by $\m$ and interact with 
the SM particles on the brane. The effects of these states 
on the SM processes, astrophysics, and cosmology on the brane
were studied in Ref. \cite {DGKN1}. It was found that 
the Hagedorn-type saturation of scattering processes on the brane
will be reached at energies of
order $E\,\sim \, \sqrt{\m\mpl}\,\sim $ TeV. 
This will manifest itself at these energies  
in production of states with the Regge spectrum spaced by $\m$.
Therefore, the present model makes a definitive predictions 
for missing energy signals is accelerator experiments in the 
TeV range \cite {DGKN1}.   
 
{\it (2) Unbroken bulk supersymmetry}. As we discussed already, 
this is something that is not
possible in any finite-volume theory.  The reason being that 
in any such a theory there is
a normalizable zero mode graviton  which must couple 
universally to all the states. 
Therefore,  in an effective low energy theory, below the compactification
scale, all the states experience non-zero Fermi-Bose mass splitting.
This is true irrespective whether they come from the brane or from the bulk.
As we discussed above, however, the infinite volume theories
allow to keep Fermi-Bose mass degeneracy  in 
the bulk while breaking SUSY on the brane.

\section{Small cosmological constant}

Let us start our discussions in this section with BPS branes. 
These branes are static even though the 4D cosmological 
constant on the worldvolume, i.e., the brane tension, takes 
a nonzero value.  For BPS branes this is achieved by tuning
the brane tension to the brane charge 
under the corresponding higher-form antisymmetric fields. 
The tuning condition is normally 
nothing but the supersymmetric BPS condition indicating 
that some part of the original bulk supersymmetry is 
(un)broken on the brane.
However, to describe the real world we would like to brake
all the supersymmetries on the brane. Hence, the tuning condition
is not going to be protected against  radiative corrections due to 
quantum loops and is expected to be violated in general.
The result is a non-BPS brane. The key question is whether a non-BPS brane 
remains static or whether it should  start inflating its
worldvolume, and  if it inflates what should be the inflation rate. 
These issues will be addressed in the next two subsections.

\subsection{Static versus inflating solutions}

In this subsection we study  the limit $V_N \to \infty$.
There are known non-BPS solutions with a nonzero tension 
the worldvolume theory of which does not posses 4D Poincare invariance
\cite {Horowitz}. However, we are after a static solution with
a 4D Poincare invariant flat worldvolume.
The existence of this solution
in the theory with the induced EH term follows from its existence
in the theory without the EH term
since ${\overline R}=0$ on the solution, by  construction. 
The solution can be parametrized as
\beq
ds^2\,=\,A^2(\rho)\,{\bar g}_{\mu\nu}(x)\,dx^\mu dx^\nu\,
-\,B^2(\rho)\,d\rho^2\,-
\,C^2(\rho)\,\rho^2\,d \Omega^2_{N-1}\,,
\label{interval}
\eeq
where $\rho \equiv \sqrt{\rho_1^2+...+\rho_N^2}$ and
the functions $A,B,C$
depend on  ${\cal E}_4$ and  $\m$.
One solution with $ {\bar g}_{\mu\nu}(x)=\eta_{\mu\nu}$,
${\overline R}({\bar g})=0$ is known  explicitly~\cite {Gregory}.
For this solution  the functions $A,B,C$
can be written as certain powers of
$\left (1- \rho_g^{N-2}/\rho^{N-2}\right )$ (see Ref. \cite {Gregory}).
If this were a unique and physical solution, being supplemented by 
the present mechanism to obtain 4D gravity on the brane, it
would solve the cosmological constant problem.
However, it 
is hard to give physical interpretation to the above solution since
it develops a naked singularity~\footnote{It is supposed to be 
a singularity at the position of the core at $\rho=0$; the latter
would be an artifact of the delta-function approximation which we use 
and could  be removed by reintroducing a small width of the  brane. 
However, in the static solution the $\rho=0$  point 
cannot be approached from the bulk since the singularity at 
$\rho=\rho_g$ prevents one to do so. 
One could cut space at $\rho=\rho_g$  and place a source at that point 
\cite {Emparan}. However, in this case the matching condition 
at $\rho=\rho_g$ require the source to have 
an equation of state which is different from that of a brane.
We would like to thanks R. Emparan for useful discussions on these
issues (see also calculations in \cite {Gregory}). 

In what follows we will argue that the physical brane inflates 
in such a way  that the singularity at $\rho = \rho_g$ is removed
and the point $\rho =0$ where the brane is located can be approached
smoothly.} in the bulk at $\rho\,=\,\rho_g$, 
where \cite {Emparan} 
\beq
\rho_g\,\sim \,\left ( {{\cal E}_4\over \m^4}  \right )^
{1\over N-2}\,{1\over \m}\,.
\label{rhog}
\eeq
In our case the value of $\rho_g$ is huge, it can be as big as 
$H_0^{-1}$. Although this solution can be used in the region $\rho>\rho_g$,
it is hard to  continue analytically to the most interesting 
for us domain $\rho < \rho_g$. 
Other solutions of this type 
can be found in Refs. \cite {Zhou}.  For these solutions 
\beq
A^2(\rho)&=&\left | { 1\,- \,(\rho_g / \rho )^{2(N-2)}  \over   
1\,+\,(\rho_g / \rho)^{(N-2)} }   \right |^{1/4}\,, \nonumber \\[3mm]
B^2(\rho)&=&\left | 1\,- \,(\rho_g / \rho)^{2(N-2)} \right |
^{1\over N-2}\times \,\left |
1\,+\,(\rho_g / \rho)^{(N-2)}    \right |^{1\over N-2}\,,
\label{AB}
\eeq
with $C^2=B^2$ and a nonzero profile of the dilaton
$$\phi \sim \left ( {\ln}| 1\,+\,(\rho_g / \rho)^{(N-2)}|
-{\ln}| 1\,-\,(\rho_g / \rho)^{(N-2)}| \right )\,.$$
As in the previous case, the solution has a naked singularity at
$\rho\,=\,\rho_g$. At this particular point the Einstein equations
are not satisfied. 
Besides this point, the manifest form of the 
above solutions have rather limited applicability 
in the present case. The reason being that 
the solutions cannot be trusted at distances 
shorter than certain $\rho_*$, $\m^{-1}\ll \rho_* \ll \rho_g$,
since the higher curvature invariants become important in that 
domain (see detailed discussions in the next section).
Because of this, and also in order 
to emphasize that our discussions below are rather general,
we will not use explicit forms of these backgrounds, but 
instead derive all the expressions for arbitrary  $A,B$ and $C$. 
We will utilize the only properties that the naked 
singularity in these solutions occurs at $\rho=\rho_g$
and that the UV physics should soften the solutions at
the scale $\rho_*$.

The singularity  at $\rho\,=\,\rho_g$  appears
because  the solutions  are highly constrained
by the requirement that the worldvolume
{\it be non-inflating} (see also discussions in section 7).
Relaxing this constraint and allowing the solution
to inflate its worldvolume,
 could smear the singularity.
A number of  examples are known  where the phenomenon of the singularity
smearing by inflation takes place.
The first example is that of a codimension one object with 
planar symmetry. Requiring that the
worldvolume is non-inflating, one obtains a solution that has
a singularity at certain distance off the core~\cite {Vilenkin1}.
This singularity is removed when the worldvolume is
allowed to inflate. In particular, the singularity is replaced by
a horizon in the bulk \cite {Sikivie}.
The distance from the core to the horizon
is determined by the curvature radius of the
worldvolume de Sitter space \cite {Sikivie}.
The next example is that of a global string
in 4D space. The constraint  that the solution is non-inflating
leads to  a naked singularity at a certain
distance from the core~\cite {CK1}.
However, an inflating solution exists
in which the singularity is replaced by the horizon~\cite {GregoryCK1}.

A similar phenomenon takes place in the case of a {\it local} string
which has the tension that is much larger than the
square of the Planck scale \cite{Cho} (this is similar
to our condition ${\cal E}_4\gg \m^4$).
Finally, a static global string in a 6D theory (a
three-brane) has a singularity in the bulk \cite {CK2}
which is removed as the worldvolume is allowed to inflate~\cite {Erich}.

Common features in all these cases
are that the inflating solution is the only nonsingular
solution. However, it is not possible
to obtain  analytic expressions
for these inflating solutions
(except for the codimension one case
of a domain wall which is not suitable for our purposes). Instead,
the proof of the existence and uniqueness of the inflating nonsingular
solution is usually given either by the methods of dynamical systems
\cite {GregoryCK1,Erich} or by numerical studies~\cite {Cho}.

Based on these facts we  expect that there exists a certain inflating
solution
in our system,
\beq
ds^2\,=\,A_1^2(t,\rho)\,dt^2\, -\, A_2^2(t,\rho)\,
{\gamma}_{ij}(x)\,dx^i dx^j\,-\,
{\gamma}_{mn}(t,\rho)\,d\rho^m d\rho^n\,,
\label{infl}
\eeq
in which  the singularity of the static solution (\ref {interval}) at
$\rho\,=\,\rho_g$ is replaced by a horizon.

It is hard to find this
solution analytically (numerical results will be discussed elsewhere).
Nevertheless, in the next subsection we will
find the properties of this solution
using the method of a probe-brane.
Here, assuming the existence
of the solution, we estimate the inflation rate of the
worldvolume using simple arguments and postponing more careful
discussions of their applicability till the next subsection.

Based on the lower-dimensional examples
discussed above, we expect that the worldvolume inflation
gives rise to a horizon in the bulk at a distance $\rho_H$ from the core
that is determined  by the de Sitter curvature radius of the 4D
worldvolume,
\beq
\rho_H~\sim~r_{\rm dS_4}~\equiv~{1\over \sqrt{\Lambda_4}}\,
\equiv \,{1\over H}\,.
\label{rhoH}
\eeq
On the other hand, we expect that this horizon is exactly what
replaces (smoothes out) the singularity at
$\rho_g$ found in the static solution. Hence, we find
\beq
\rho_H\,\sim \rho_g \,\sim \, \left ( {{\cal E}_4\over \m^4}  \right )^
{1\over N-2}~{1\over \m}\,.
\label{rhoH1}
\eeq
Substituting this into Eq.~(\ref {rhoH})
we obtain
\beq
H\,\sim\,\m\,\left (  {\m^4 \over {\cal E}_4  } \right )^{1\over N-2}\,.
\label{HNr}
\eeq

\vspace{1mm}

We would like to make a few comments here.
The simple arguments presented above 
do not use the fact that there is an induced EH
term on the brane and that there are 
higher derivative terms in the bulk. However, 
as we will see in the next sections, the existence 
of these terms is absolutely crucial for the above naive 
arguments to hold.
Moreover, in general one  should  
also include the effect of the induced EH term in Eqs. (\ref {rhoH}, 
\ref {rhoH1}). In that case we  should make the following
substitution  in these equations:
${\cal E}_4 \to {\cal E}_4 -\mpl^2 H^2$. 
This gives rise to a second solution (on top of 
the one in (\ref {HNr}))  with the inflation rate 
$H^2\sim {\cal E}_4 /\mpl^2$. 
In this solution, the tension curves strongly the
4D worldvolume, but affect only mildly the bulk space.
Therefore, the effect is opposite to the one obtained in the 
solution (\ref {HNr}) considered in the present work. 
A special interesting case of this is a possibility to have
``stealth branes'' \cite {VC,Dick} for which 
${\cal E}_4 -\mpl^2 H^2$ is zero. 
On these branes the induces EH term completely neutralizes 
the brane tensions, so that the brane has no gravitational
effect of the outside world. Hence, such a brane would be
invisible for a bulk observer.  
These branes can be produced in a tunneling process  
via bubble nucleation \cite {VC}. Hence, 
there are three different  solutions.  
These solutions are distinguished by 
initial and boundary conditions \cite {VC,Dick}. 
In the present context we consider the 
initial and boundary conditions on the brane 
that are relevant for the solutions of the type 
(\ref {HNr}).

After these comments let us discuss 
an explicit example of a codimension one object, 
with the known  exact solution ~\cite {Vilenkin1,Sikivie}.
For this solution ~\cite {Vilenkin1,Sikivie}
$\rho_H\sim 1/H$.
Hence, we reproduce the known result \cite {Cedric}
 $$H\,\sim\,{{\cal E}_4 / \m^{3} }\,.$$
Therefore, as in four
dimensions, the inflation rate
in a  5D theory grows when ${\cal E}_4$ is increasing.
This is a property of $N=0$ and $N=1$ theories.
However, for $N>2$ the exponent on the right-hand side
 of Eq.~(\ref {HNr})
is positive, and we get a flipped relation
where the inflation rate reduces
as ${\cal E}_4$ increases.

\subsection{The probe-brane method}

Let us go back to a static solution
discussed in the previous section, that
has a singularity at $\rho_g$. Below we are going to use
this singular solution in order to argue that
a nonsingular inflating solution should also exist
and to estimate its rate of inflation.
For this we use the method of a probe-brane.
Let us describe this procedure in detail.
The singularity at $\rho = \rho_g$ should be expected 
to be smoothed out in a physical solution. That is to say,
the Einstein equations for the physical solutions should be 
satisfied everywhere including the point $\rho = \rho_g$.
However, whatever mechanism smoothes out this singularity
the very same mechanism can give rise to an additional contribution
to the energy density and make the original static solution
time dependent \cite {Niles}. The major goal in that case is to find out
how strong is the time dependence of the solution,
in particular, what is the inflation rate on the worldvolume.
For this we can use the probe-brane method.

Let us compactify  extra  space at a huge distance 
$l=\rho_g$. This is only possible if we place a probe (anti)brane 
at the point $l=\rho_g$. 
The compactification radius is huge, of the order of
the Hubble size. Nevertheless,  
the role of the probe-brane is 
to enforce that the Einstein equations are satisfied
everywhere including the point  $l=\rho_g$.
As we  argue below, for a generic value of the 
probe brane tension this procedure deforms the original 
static solution so that it will start to inflate.

To start with let us neglect the bulk higher derivative terms.
The role of those terms, as we will discuss in the next section, 
is to make the solution sensible in a region close to the brane core. 
The relevant part of (\ref {actD}) takes the form
\beq
{\cal L}_{4D}\,=\,
\m^{2+N}\,\int_0^l\,d^N\rho\,\sqrt{G}\,{\cal R}(G)\,+\,
\sqrt{\bar g}\,\left (\mpl^2\,  {\overline R}(\bar g )\,+\,
\,{\cal E}^{\prime}_4 \right )\,+\,\int_0^l\,d^N\rho\,\sqrt{G}
\,{\cal E}_c[l] \,,
\label{actc}
\eeq
where 
\beq
{\cal E}^{\prime}_4 \,\equiv \, {\cal E}_4\,+\, T_{\rm probe}\,,
\label{probe}
\eeq
denotes the sum of the brane tension ${\cal E}_4$
and the tension of a probe brane $T_{\rm probe}$.
Furthermore, $ {\cal E}_c[l]$ denotes the Casimir energy. The latter
arises~\footnote{
If SUSY is broken on the brane but not in the bulk, the SUSY breaking
will penetrate  to the bulk as long as $l<\infty$.
This will give rise to a contribution to the vacuum energy
neglected above which for large $l$ is suppressed as
$\sim ({\cal E}_4/\m^{2+N}l^N)\m^N$.}
because of the compactness of extra space. It is clear that
$ {\cal E}_c[l]\,\sim\,1/l^{4+N}$ and at large $l$ it vanishes.

Since the extra space is terminated at a finite proper distance $l$,
the spectrum of gravitons consists of a massless
zero mode and a tower of  massive
KK states with the masses $\sim 1/l$.
Let us consider dynamics  at very low energies $E$
that are relevant for the cosmological
constant problem. Naively, for $E\le l^{-1}$
all the massive KK modes are decoupled,
only the zero mode remains. However,
this statement needs a more careful justification
since its validity is limited, as we shall see now.

Consider an observer who sits in a brane core in the vicinity of the point
$\rho \sim 0$ and is able to probe distances that are smaller
than the brane thickness $\Delta $.
This observer finds him/herself in a $(4+N)$-dimensional
space that has a uniform energy density
${\cal E}_4/\Delta^{N}$. If the $(4+N)$-dimensional
Hubble distance of this  observer, $H^{-1}_W$,
is smaller than the  brane width $\Delta$,
the observer would find him/herself in a $(4+N)$-dimensional
space with  positive energy density ${\cal E}_4/\Delta^{N}$
that should inflate along all  $(4+N)$-dimensions (in
analogy with the topological inflation of Refs.~\cite {LindeT,VilenkinT}).
This would invalidate
our arguments on the low-energy description.
In order for this not to happen, the $(4+N)$-dimensional
Hubble distance $H^{-1}_W$  should be larger than  the brane thickness
$\Delta$.  This puts a bound on ${\cal E}_4$.
Let us find this bound. A vital role here is played by the
induced EH term.  Inside the brane core  gravity
is weak precisely because of this term, and the Hubble distance
$H^{-1}_W$ becomes large.
Although the induced term
cannot be written as a local term when  the $\Delta\ne 0$ effects are
kept \cite {DGHS}, nevertheless, one can think of the induced term as a
renormalization of the  bulk EH term inside the brane core of
the size $\Delta$. Hence, $H^{-1}_W$ can be estimated from the following
expression $${\mpl^{2}\over \Delta^N}\,H^{2}_W \, 
\sim \,{{\cal E}_4 \over \Delta^{N}}\,. $$
This gives rise to the bound
$${\cal E}_4  \ll \mpl^2/\Delta^2\,.$$

A typical value of the brane thickness, as was discussed in Sect. 3,
is $\Delta \sim \msm^{-1}$.
Hence, we see that the topological inflation does not take place
even for the energy  density as large as $\msm^4$. This is all we want.
If we were to ignore the induced term on the brane, we would find that
for a  brane the tension of which exceeds ${\rm TeV}^4$
the topological inflation would take place unless the brane width was
 several orders of magnitude smaller than the Planck length $\mpl^{-1}$.

After this digression, let us proceed with a brane tension which satisfies
the above constraint. We  set $l$
to be somewhat smaller than the Hubble size,  
and turn to the distances of the order of $H_0^{-1}$. At those distances
all massive KK modes are decoupled.
The low energy effective Lagrangian for the zero mode reads
\beq
{\cal L}_{\rm eff}\,=\,
(\m^{2+N}\,l^N\,+\,\mpl^2)\,\sqrt{{\bar g}_{\rm zm}}\,{\overline R}\,
({\bar g}_{\rm zm})
\,+\,\sqrt{{\bar g}_{\rm zm}}\,\left (  {\cal E}^{\prime}_4\,+\,l^N\,
{\cal E}_c[l]\right )
\,,
\label{lagr4}
\eeq
where ${\bar g}_{\rm zm}$ denotes the zero mode.

The Einstein equations for this Lagrangian density
yield a solution with the following  inflation rate:
\beq
H^2\,\equiv \,\Lambda_4 \,\sim \,{{\cal E}^{\prime}_4\,+\,l^N\,{\cal E}_c[l]
\over \m^{2+N}l^N \,+\,\mpl^2  }\,.
\label{Hcc}
\eeq
Hence, we see that a solution will inflate in general.
One could certainly choose  the tension of a probe brane
in (\ref {probe}) such that  
$H$ vanishes in (\ref {Hcc}). However, 
this cannot be  considered as a solution to the cosmological 
constant problem as it is nothing but another {\it fine tuning} of 
the parameters of the  theory. Therefore, in what follows 
we abandon this possibility  and consider a case when the 
tension of a probe brane takes
a {\it generic} value not necessarily the one 
that nullifies $H$. The magnitude of  
$T_{\rm probe}$ can generically be of the order of ${\cal E}_4$
but otherwise bigger (or smaller) by some factor.
From the technical point of view that means that 
in (\ref {Hcc}) we can think that ${\cal E}^{\prime}_4$ takes the 
value  that is within  an  order of magnitude of 
the brane tension ${\cal E}_4$. 

If we were confined to a theory in which the
conventional KK relation (\ref {KK}) holds
we would obtain an unacceptable result from (\ref {Hcc})
 $$\Lambda_4 \,\sim \, ( {\cal E}^{\prime}_4\,+\,l^N\,{\cal E}_c)/
\mpl^2\,\gsim \, (10^{-3}~{\rm eV})^2\,.$$
However, in the present model the relation
(\ref {KK}) does not hold, as we have emphasized in
(\ref {antiKK}). Therefore, we could (and we should) 
consider a regime where
\beq
\m^{2+N}l^N \,=\,\m^{2+N}\rho_g^N~\gg~\mpl^2~.
\label{antikk}
\eeq
In this regime, the second terms in the numerator and denominator
of Eq.~(\ref {Hcc}) can be neglected compared to the first terms,
respectively.
Furthermore,  the value of $\Lambda_4$ can be made much smaller then
$(10^{-3}~{\rm eV})^2$ by making $l$ large.

How large could $l$ be?
Since we  terminate  extra space at the singularity of the original
static solution, $l\,=\, \rho_g\,.$
On the other hand, in order for this consideration to be relevant
for the cosmological constant problem,
$l$ should be somewhat smaller than the present-day Hubble size,
$$l \lsim H_0^{-1}\,.$$
Hence, we get an estimate for the rate of  inflation,
\beq
H^2\,\sim \,{{\cal E}^{\prime}_4\,
\over \m^{2+N}\,\rho_g^N}\,.
\label{hor}
\eeq
Substituting into Eq. (\ref {hor}) the expression
$$\rho_g \sim \m^{-1}({\cal E}^{\prime}_4/\m^4  )^{1/(N-2)}$$
 we find the same equation (\ref {HNr})
from which we see that the inflation rate $H$ decreases
with increasing  ${\cal E}^{\prime}_4$ if  $N>2$.

Note that the probe-brane  and compactification
is used in this subsection as technical tools to 
argue that the truly physical solution in uncompactified
space will inflate with the rate (\ref {HNr}). Certainly, if the latter
solution is found by some other means, such as, e.g., 
numerical simulations, the probe-brane and compactification method 
will just become redundancy.
  
Finally, as representative
examples, we give estimates for $H$
as determined from (\ref {HNr}),
\beq
H\,\sim \,10^{-33}~{\rm eV}~~{\rm for}~~N=4,~~\m\,\sim \,10^{-3}~{\rm
eV}~,
~~{\cal E}^{\prime}_4 \,\sim \,\left ({\rm TeV}\right )^4~;\nonumber \\[0.3cm]
H\,\sim \,10^{-33}~{\rm eV}~~{\rm for}~~N=6,~~\m\,\sim \,10^{-3}~{\rm
eV}~,
~~{\cal E}^{\prime}_4 \,\sim \,\left (\,\mpl \,\right )^4\,.
\label{etimates}
\eeq
These values are consistent with experimental data.

\section{Softening of  the bulk  metric}

 In this section we shall study the behavior of the background solution
near the brane. As we shall see, the 
understanding of this behavior is crucial for determining
the crossover scale at which gravity on 
the brane switches from the 4D regime 
to a higher-dimensional one.

As we discussed above, the static solution of the type 
(\ref{interval}) could in general have  at least 
two types of singularities. One is a
naked singularity at $\rho = \rho_g$. As was mentioned above,
this is an artifact of over-constraining the system at hand, by requiring 
the  flatness of the 4D metric. This singularity can be smoothed out by 
relaxing the requirement of the flatness and allowing for an inflating
brane. 

The second type of singularity occurs at $\rho = 0$ and is not
related to inflation, but rather to the excessive warping near the origin.
Smoothing out the brane (by substituting
the delta function brane by a smooth sharply localized function
of a finite width) would cut-off infinities. Nevertheless, 
the curvature of the space-time near the brane would still
be much higher than the fundamental scale $M_*$ (i.e. the string scale)
and an effective field theory approximation would break down.
We would like to analyze the later issue in detail in the present section.
Before we do so, let us recall that 
in the previous sections we argued that the physical solution is the 
one that inflates its worldvolume. However, the inflation rate
is tiny. Therefore, for practical calculations it is 
easier to neglect this weak time-dependence of the metric  
and to use the static solution (dealing carefully with 
the singularities). The presence of a small inflation rate
is not going to affect any  predictions for distances bigger than 
$0.1$ mm and smaller that  the present day Hubble size.

Consider, for instance,  the behavior of the Riemann tensor 
${\cal R}_{ABCD}$ in such a background. The bilinear curvature invariant
constructed out of this tensor scales as (near the core)
\beq
 {\cal R}_{ABCD}{\cal R}^{ABCD} \sim {\rho_g^n \over \rho^{n+4}} \,,
\label{38}
\eeq
where $n$ depends on the precise form of the solution.
For the branes of interest the right-hand side in Eq. (\ref{38})
 becomes larger
than
$M_*^4$ at a macroscopic distance from the core 
$\rho=\rho_* \gg M_*^{-1}$. This
would mean, in particular,  that at distances $\rho\lsim \rho_* $ 
a freely-falling observer would experience a tidal force
exceeding,  by many orders of magnitude,  the fundamental scale,
and, thus, the solution cannot be trusted below $\rho_*$.
(Note that because of the nontrivial geometry $\rho_*$ is not a proper
distance. The invariant statement is that the solution breaks down when
the curvatures becomes of the order of $M_*^2$.)
The question is how can we
understand the required smoothness of the solution near the brane from
an effective field theory perspective?

The problem that we encounter here is by no means  
different than that  emerging in a problem of  
a freely-falling observer
in the vicinity of an ordinary four-dimensional macroscopic Schwarzschild
black hole. At distances 
\begin{equation}
\rho_* \sim M_P^{-1} \left ({M\over M_P}\right )^{1/3} \gg M_P^{-1},
\end{equation}
the standard Schwarzschild 
solution obtained for a point-like object should break down
\footnote {Modulo the fact that in the interior of the 
Schwarzschild solution space and time interchange.}.

In both cases above  one can restore the validity
of  the solutions  only 
by taking  account of higher derivative operators (HDO's) coming from the
underlying fundamental theory of gravity. These correction cannot
be neglected at distances $\sim \rho_*$. Thus, one has to take into account
all possible higher-dimensional invariants suppressed by the scale $\m$
in the bulk action, 
\beq
{S}_{\rm bulk}\,=\,
\m^{N+2}\,\int\,d^4x\,d^N\rho\,\sqrt G\, \left\{ {\cal R}\,+
\frac{\alpha}{\m^{2 }}\,{\cal R}_{ABCD} {\cal R}^{ABCD}\, + \,...\right\}\,,
\label{bulknew}
\eeq
(similar terms
added to the brane action are suppressed by powers of $\mpl$ and are
less important).
Here $\alpha$ is a numerical coefficient of order one.
These terms modify the Einstein  equations in the following way:
\begin{eqnarray}
&&\m^{2+N}\,\left [{\cal R}_{AB}\,-\,{1\over 2}G_{AB}\,\left({\cal R}\, + 
 \frac{\alpha}{\m^{2}}{\cal R}_{CDMN}\,{\cal R}^{CDMN}\,\right)\right.
\nonumber\\[3mm]
&&+ \left. \frac{\alpha}{\m^{2}}\,\left (\, 2{\cal R}_{ACDM}{\cal R}^{CDM}_B\,
+  4\nabla^C\nabla^D\,{\cal R}_{ACBD}\right)\right ]\, + \,...
\nonumber\\[3mm]
 &&+\, \mpl^2\,\delta^\mu_A \delta^\nu_B \, \d \,
\left ({\overline R}_{\mu\nu}\,-\,{1\over 2}{\bar g}
_{\mu\nu}\,{\overline R} \right )\,\nonumber\\[0.3cm] && =\,
 {\cal E}_4 \,{\bar g}_{\mu\nu}\,\delta^\mu_A \delta^\nu_B \d \,.
\label{eeq11}
\end{eqnarray}

Strictly speaking, one has to include an infinite number of 
higher-dimensional curvature invariants, and, therefore, finding the exact
form of the solution is impossible. Fortunately, the precise form of the
metric near the
brane is not
important for us. What is crucial, however, is the softening of the background
metric. The fact that inclusion of higher-derivative terms 
{\em must} lead to the 
softening can be explicitly seen in the linearized example (note that
linearization {\it per se} does not make the problem milder.
In the Einstein gravity without the higher-curvature terms the 
ultraviolet
singularity in the linearized solution is as severe as in the non-linear one).

For the purpose of a simple and transparent illustration we shall ignore
the tensorial structure and consider the linearized scalar gravity. The
equation then becomes 
\beq
 \left ( \m^{2+N}\,\partial_A\partial^A \,
\,
 +\, \mpl^2 \, \d \,
\partial_{\mu}\partial^{\mu}\right )\,\Phi   = 
 {\cal E}_4 \,\d \,.
\label{scalargr}
\eeq
Here $\Phi$ should be thought of as a counterpart of the 
classical background. The solution independent of 
the world-volume space-time coordinates is  
a massless Euclidean Green's function in the transverse space
\beq
\Phi(\rho) \,= \, 
{{\cal E}_4 \over \m^{2+N}}{1\over \rho^{N-2}}\,.
\eeq
It is obviously singular in the ultraviolet. The singularity is smoothed out by
adding  higher-derivative
terms in the action. The point can be illustrated by adding a single
additional power
of the operator $\partial_A\partial^A$ in the left-hand side of
Eq.~(\ref{scalargr}),
\beq
 \m^{2+N}\,\left (\,\partial_A\partial^A \, + \,
{(\partial_A\partial^A)^k\over \m^{k-2}} \right )\Phi
\,+
\, \mpl^2 \, \d \,
\left (\partial_{\mu}\partial^{\mu}\, \Phi \right )  = 
 {\cal E}_4 \,\d \,.
\label{scalargr1}
\eeq
The solution to this equation,
\beq
\Phi(\rho)\,=\,{{\cal E}_4 \over \m^{2+N}}\,
\int \frac{{d}^{N}q}{\left( 2\pi \right) ^{N}}
\, \frac{e^{iq\rho}}{\,q^{2}\,+ \,
\left(q^{2}\right)^{k}\m^{ 2-k} } \,,
\eeq
is non-singular at $\rho = 0$ for $2k> N+ 1$ and $ N\geq 2 $.
In fact, $\Phi(0)\,\propto \,{{\cal E}_4 /\m^{4}}\,.$
Therefore, based on this example, we expect that 
HDO's give rise to UV softness of the classical backgrounds
at distances $\rho \lsim \rho_*$. This feature is expected
to be true in a theory of quantum gravity.

\section{Four-dimensional gravity on a brane}

In this section we study whether the laws of 4D gravity are reproduced
on a brane at observable distances. The 4D laws are 
certainly obtained when ${\cal E}_4=0$,
as it was shown in Refs. \cite {DGP,DG,DGKN1,DGHS}.
The crucial difference here is the presence of a non-zero brane
tension  which gives rise to new nontrivial properties of the 
classical background. For $N=2$ case this was studied
in Ref. \cite {zura2} with the conclusion that the properties
obtained in Refs. \cite {DGP,DG,DGKN1,DGHS} hold unchanged.
In the next two subsections  
we will perform the analysis for $N>2$ in terms of the Green's 
function on the brane following Refs.  \cite {DG,DGHS},  
as well as in terms of the KK modes following the method
of Ref. \cite {DGKN1}. We will show that at observable distances
the 4D laws of gravity are indeed reproduced. 
As before, we will neglect a tiny time dependence
of the classical background and will treat it as static.

\subsection{Propagator analysis}

The nature of gravity on the brane perhaps is simpler understood from
the propagator analysis. The equation for the graviton two-point
Green's function (we omit tensorial structures) 
takes the following form
\beq
\m^{2+N}{\hat {\cal O}}_{4+N}\,G(x,\rho)\,+\,{\mpl^2\,\delta
(\rho)
\over \rho^{N-1}\,A^2(\rho)}\,
{\hat {\cal O}}_4\,G (x,0)\,= T\,\delta^{(4)}(x)\,{\delta
(\rho)\over \rho^{N-1}}\,,
\label{Prop}
\eeq
where $T$ denotes the source (which will be put equal to 1 below) and 
\beq
{\hat {\cal O}}_{4+N}\,\equiv
\,{1\over \sqrt{G}}\partial_A\,\sqrt{G}\,G^{AB}\,
\partial_B\,+{\rm higher~~derivatives},~~~~~~~
{\hat {\cal O}}_4\, \equiv  \,\partial^\mu\partial_\mu \,.
\label{B0}
\eeq
Using the technique of Ref. \cite{DG}, the scalar part of the solution in the
Euclidean four-momentum space can be written in the form\footnote{
Note that in the warped case the scale $M_{\rm ind}$ differs from that 
of the flat case by a constant multiplier $A^2(\Delta)$. For simplicity 
this won't be depicted manifestly below.}
\beq
G(p,\rho )\,= \,{D(p,\rho) \over\mpl^2 p^2 D(p, 0) \,+\, \m^{2+N}}\,,
\eeq
where $D(p,\rho)$ is the Euclidean 4-momentum Green's function of the
bulk operator ${\hat {\cal O}}_{4 + N}$, that is 
${\hat {\cal O}}_{4 + N} D(p,\rho)=\delta(\rho)/\rho^{N-1}$. 
What is crucial for us is the behavior of the 
Green's function on the brane
\beq
G(p,\rho=0 )\,=\,{1 \over \mpl^2 p^2\,+\,\m^{2+N} D^{-1}(p, 0)}\,.
\label{gee}
\eeq
Let us discuss this expression first. The denominator in 
(\ref {gee}) consists of two terms. The first term, 
$\mpl^2 p^2$, is what gives rise to 4D behavior. 
The second term in the denominator, $\m^{2+N} D^{-1}(p, 0)$, sets the 
deviation from the 4D laws and is due to the infinite-volume extra bulk.
Therefore, in the regime when $\mpl^2 p^2$ dominates over 
$\m^{2+N} D^{-1}(p, 0)$ we get 4D laws, while in the opposite case
we obtain the higher-dimensional behavior. The question is what is
the crossover scale at which this transition occurs.
To answer this question we need to know
the expression for $D(p,0)$. Let us start for simplicity
with the case when ${\cal E}_4=0$, i.e.,  
the background metric is flat. 
We will denote the corresponding Green's function 
by $D_0(p, \rho)$ to distinguish it from $D(p,\rho)$.
Moreover, let us drop for a moment
higher-derivatives in the expression for 
${\hat {\cal O}}_{4 + N}$.  In this case 
$D_0(p,\rho)$ is nothing but the Green's function of 
the  $(4+N)$-dimensional d'Alambertian. Its behavior at 
the origin is well known:
\beq
D_0(p,\rho \to 0)\,\sim \,{1\over \rho^{N-2}}\,.
\label{dal}
\eeq  
Hence, $D_0(p,0)$ diverges and therefore the term 
$\m^{2+N} D^{-1}(p, 0)=\m^{2+N} D_0^{-1}(p, 0) $
in Eq. (\ref {gee}) goes to zero. This would indicate that
4D gravity is reproduced on the brane at all distances. However, 
the UV divergence in (\ref {dal}) is unphysical. 
This divergence is smoothed out by UV physics 
\cite {DG,Wagner,Kiritsis,DGHS}. 
In reality the bulk action and the operator 
${\hat {\cal O}}_{4 +N}$ contain an infinite number of
high-derivative terms that should smooth out singularities in the 
Green's function in  (\ref {dal}).
Since these HDO's are suppressed by the scale $\m$, it is natural that
the expressions (\ref {dal}) is softened at the very same scale 
$\rho \sim \m^{-1}$. As a result one obtains \cite {Kiritsis,DGHS}
$D_0(p,\rho=0)\sim \m^{N-2} (1+{\cal O} (p/\m))$. 
Substituting the latter expression
into (\ref {gee}) we find that $\m^{2+N} D^{-1}(p, 0)= \m^{2+N} 
D_0^{-1}(p, 0)\sim \m^4 $.
Therefore the crossover scale is $r_c\sim \mpl /\m^2 \sim 10^{28}$ cm.
At distances shorter than this 4D laws dominate.

Let us now switch on the effects of a non-zero tension ${\cal E}_4$. 
The background in this case is highly distorted. 
The distortion is especially strong near the brane.
Let us start again with the case when the HDO's are neglected and 
${\hat {\cal O}}_{4 +N}$ contains only two derivatives at most.
Then, the expression for the $D$-function takes the form:
\beq
D(p,\rho)\,\sim\,D_0(p,\rho)\,{\cal F}(\rho_g / \rho )~,
\label{solsol}
\eeq  
where, as before, $D_0(p,\rho) \sim {1/\rho^{N-2}} $  and 
${\cal F}$ is some function which is completely determined by 
the background metric (by the functions $A,B$ and $C$)
and ${\cal F}(0)={\it const}$.  In the region where the solution 
of the Einstein equations can be trusted, 
${\cal F}$ can be approximated as follows,
$ {\cal F}(\rho_g / \rho)=( {\rho_g / \rho} )^{\alpha^2}+c$,
where $\alpha$ and $c$ are some constants determined by $N$.
If we were to trust this solution all the way down to the point 
$\rho =0$ we would obtain again that  $\m^{2+N} D^{-1}(p, 0)=0$ and 
that gravity is always four-dimensional on the brane.
However, as we discussed above (see also the previous section), 
the  existence of high-derivative terms tells us that the 
background solution cannot be trusted for distances
$\rho \ll \rho_*$. 
In general, $\rho_*=\m^{-1}(\rho_g \m)^{\gamma}$ with 
$ \gamma \ll 1$ and  $\rho_* \lsim  \rho_g$.
Thus, for $\rho \ll \rho_*$  the higher curvature 
invariants become large in units of $\m$ and infinite number of them should 
be taken into account. In order to find the effect of this softening,
let us take a closer look at the expression (\ref {solsol}).
There are two sources of singularities in this expression. 
The first one  emerges on the r.h.s. of (\ref {solsol})
as a multiplier, $D_0\sim {1/ \rho^{N-2}}$; this  singularity 
was discussed above in (\ref {dal})
and is independent of the background geometry.
Instead, it emerges when the operator 
${\hat {\cal O}}_{4 +N}$ is restricted to the quadratic order only.
We expect that this singularity, as before, is softened at the 
scale $\m^{-1}$ after the higher derivatives are 
introduced in ${\hat {\cal O}}_{4 +N}$. Hence, in (\ref {solsol})
when we take the limit $\rho \to 0$ 
we should make a substitution $D_0\sim {1/ \rho^{N-2}} \to \m^{N-2}
(1+{\cal O} (p/\m))$.
On the other hand, the second source of singularity in 
(\ref {solsol}) is due to the function  ${\cal F}$.
This singularity is directly related to the fact that 
the background solution breaks down at distances of the order of 
$\rho_*$. As we discussed in the previous section, 
the UV completion of the theory by HDO's should smooth out 
this singularity in the background solution.
In order to get the crossover scale
we can use the following procedure which overestimates
the value of   $\m^{2+N} D^{-1}(p, 0)$.
In the limit $\rho \to 0$ we could make the following substitution in
the expression for ${\cal F}$ and in
(\ref {solsol}), $ \rho_g/\rho \to \rho_g/\rho_*$.
Using these arguments we find $ D(p, 0)\lsim \m^{N-2} [
({\rho_g / \rho_*})^{\alpha^2}+c ]$. Moreover, taking into account that
$\rho_* \lsim \rho_g$ we get: $\m^{2+N} D^{-1}(p, 0)\lsim \m^{4}$.
Therefore, we conclude that, as in the zero-tension case, 
the crossover distance\footnote{If we were to assume  that the background 
solution is softened at $\m^{-1}$ rather than at $\rho_*$,
we would obtain even larger value for the 
crossover scale. This can be turned around to make  the following 
observation. If the background metric softens at $\m^{-1}$, and/or 
if $\rho_g\gg \rho_*$, the value of $\m$ should not necessarily 
be restricted to $10^{-3}$ eV, but  can be much higher. 
Unfortunately, these properties does not seem to allow 
further analytic investigation.} in the non-zero tension  
case can be of the order of $10^{28}$ cm.

\subsection{KK mode analysis}

The purpose of this section is to study the
effect of a nonzero brane tension on
4D gravity in terms of the KK modes.
The Einstein equations (with up to two derivatives)  
that follow from the action (\ref {actD}) take the form
\beq
&&\m^{2+N}\,\left ({\cal R}_{AB}\,-\,{1\over 2}G_{AB}\,{\cal R}\right )
\, +\, \mpl^2\,\delta^\mu_A \delta^\nu_B \, \d \,
\left ({\overline R}_{\mu\nu}\,-\,{1\over 2}{\bar g}
_{\mu\nu}\,{\overline R} \right )\,\nonumber\\[0.3cm] && =\,
 {\cal E}_4 \,{\bar g}_{\mu\nu}\,\delta^\mu_A \delta^\nu_B \d \,.
\label{eeq1}
\eeq
Below we consider fluctuations $h_{\mu\nu}(x,\rho_n)$ that are relevant
for
4D interactions on the brane,
\beq
ds^2\,=\,A^2(\rho)\,\left [\eta_{\mu\nu}\,+\,h_{\mu\nu}(x,\rho_n)\right]
\,dx^\mu dx^\nu\,-\,B^2(\rho)\,d\rho^2\,-
\,C^2(\rho)\,\rho^2\,d \Omega^2_{N-1}\,.
\label{fluct1}
\eeq
In what follows we use  the transverse-traceless gauge
$$\partial^\mu h_{\mu\nu}=0=h^{\alpha}_{\alpha}\,.$$
As  typically happens in warped backgrounds,
equations for graviton fluctuations
are identical to those for a minimally coupled
scalar~\cite {Borut,Csaki}. The present case is no
exception.  Equation (\ref {eeq1}) on the
background defined in Eq.~(\ref {fluct1}) takes the form
\beq
\m^{2+N}{\hat {\cal O}}_{4+N}\,h_{\mu\nu}(x,\rho_n)\,+\,{\mpl^2\,\delta
(\rho)
\over \rho^{N-1}\,A^2(\rho)}\,
{\hat {\cal O}}_4\,h_{\mu\nu}(x,0)\,=\,0\,,
\label{BBox}
\eeq
where
\beq
{\hat {\cal O}}_{4+N}\,\equiv
\,{1\over \sqrt{G}}\partial_A\,\sqrt{G}\,G^{AB}\,
\partial_B\,,~~~~~~~
{\hat {\cal O}}_4\, \equiv  \,\partial^\mu\partial_\mu \,.
\label{B}
\eeq
To simplify  Eq.~(\ref {BBox}) we
turn to spherical coordinates with respect to $\rho_n, n=1,2,.., N$,
and decompose fluctuations as follows:
\beq
h_{\mu\nu}(x,\rho)\,\equiv \, \epsilon_{\mu\nu}(x)\,\sigma(\rho)\,
\phi(\Omega)\,,
\label{fluc}
\eeq
where the components in Eq.~(\ref {fluc}) satisfy
the conditions
\beq
{\hat {\cal O}}_{4+N}\,\epsilon_{\mu\nu}(x)\,& = &\,{1\over A^2}\,
\partial^\mu\partial_\mu
\,\epsilon_{\mu\nu}(x)\,=\,-{m^2\over A^2}\,\epsilon_{\mu\nu}(x)\,,
\label{eps} \\[0.3cm]
{\hat {\cal O}}_{4+N}\,\phi(\Omega)\,& = &
{l\,(l+N-2)\over C^2\,\rho^2}\,\phi(\Omega)\,.
\label{phi}
\eeq
Using these expressions we rearrange Eq.~(\ref {BBox})
as follows:
\beq
\left \{ {1\over \sqrt{G}}\partial_\rho\,\sqrt{G}\,G^{\rho \rho}\,
\partial_\rho\,+\,{l\,(l+N-2)\over C^2\,\rho^2}\,
-\,{m^2\,\mpl^2\,\delta (\rho)
\over \m^{2+N}\,\rho^{N-1}\,A^2(\rho)}
\right \}\,\sigma\,=\,
{m^2\over A^2}\,\sigma\,.
\label{eq}
\eeq
Our goal is to rewrite  this expression
in the form of a Schr\"odinger equation
for fluctuations of mass $m$.
We follow the method of Refs.~\cite {CK2,Emparan}.
It is useful to introduce a new function
\beq
\chi\,=\,{G^{1/4}\over \sqrt{A\,B}}\,\sigma\,,
\label{chi}
\eeq
and a new coordinate
\beq
u\,\equiv\,\int_0^{\rho}\,d\tau \,{B(\tau)\over A(\tau)}\,.
\label{u}
\eeq
In terms of these variables Eq.~(\ref {BBox})
takes the form
\beq
\left \{
-{d^2\over du^2}\,+\,V_{\rm eff}(u)
\,+\,{A^2\,l\,(l+N-2)\over C^2\,\rho^2}\,
-\,{m^2\,\mpl^2\,\delta (\rho)
\over \m^{2+N}\,\rho^{N-1}}
\right \}\,\chi\,=\,
{m^2}\,\chi\,,
\label{equ}
\eeq
where the effective potential $V_{\rm eff}(u)$
is defined as
\beq
V_{\rm eff}(u)\,=\,{\sqrt{A\,B}\over G^{1/4} } \,
{d^2\over du^2}\,\left ( {G^{1/4}\over \sqrt{A\,B}}
\right )\,.
\label{Veff}
\eeq
Note that the first two terms in (\ref {equ}) can be
rewritten as follows:
\beq
-{d^2\over du^2}\,+\,V_{\rm eff}(u) \, = \, \left ({d\over du}\,+\,
 {d {\cal B} \over du}  \right )\, \left (-{d\over du}\,+\,
 {d {\cal B} \over du}  \right )\,,
\label{QQ}
\eeq
where 
\beq
{\rm exp} \left ( {\cal B} \right )\, \equiv\, 
{G^{1/4}\over \sqrt{A\,B}}\,.
\label{B1}
\eeq
With an appropriate physical boundary conditions 
the operator on the r.h.s. of (\ref {QQ})
is self-adjoint positive-semidefinite with 
a complete set of eigenfunctions of 
non-negative eigenvalues.

\vspace{0.1cm}

Let us analyze Eq.~(\ref {equ}), in particular the
properties of the KK modes that follow from it.
What is crucial  for our purposes is
the value of the KK wave functions on the brane, i.e.,
$|\chi (m,\rho=0)|^2$.
The later determines a potential between two static
sources on the brane \cite {DGKN1}.
We would like to compare the
properties of $|\chi (m,\rho=0)|^2$ which are known~\cite {DGKN1,DGHS}
 only for ${\cal E}_4\,=\,0$,
with the properties obtained at ${\cal E}_4\ne 0$.

First, we recall  the properties
of $|\chi (m,\rho=0)|^2$
for $N=1$ and tensionless brane, ${\cal E}_4=0$.
In this case $A=B=1$, $C=0$ and $l=0$. Hence, $u=\rho$ and
$V_{\rm eff}(u)=0$. Equation (\ref {equ}) takes the form
\beq
\left \{
-{d^2\over d\rho^2}\,
-\,{m^2\,\mpl^2
\over \m^{3}}\, \delta (\rho)
\right \}\,\chi\,=\,
{m^2}\,\chi\,.
\label{redeq}
\eeq

\vspace{0.1cm}

For each KK mode of mass $m$ there is a delta-function
{\it attractive} potential, the strength of which is proportional
to the mass of the mode itself.
Hence, the higher the mass, the more the influence of the potential is.
The attractive potential leads to a suppression of
the wave function at the origin (suppression of $|\chi (m,\rho=0)|^2$).
Therefore, the larger the mass of a KK state,
the more  suppressed  is its wave-function at zero.

Simple calculations in this case yield
$$|\chi (m,\rho=0)|^2={4/(4+m^2r_c^2})\,,$$
where $r_c \sim \mpl^2/\m^3$.
This should be contrasted with the expression for $|\chi (m,\rho=0)|^2$
in a theory with no brane induced term
(i.e., with no potential in Eq.~(\ref {redeq})). In that case 
$|\chi (m,\rho=0)|^2 \,=\,1$.
We see  that the KK modes with masses $m\gg r_c^{-1}$ are
suppressed on the brane. The laws of gravity on the
brane are provided  by light modes with  $m\,\lsim\, r_c^{-1}$.
This warrants~ \cite {DGP,DGKN1} that at distances
$r\lsim  r_c$ measured along the brane
the gravity laws  are
four-dimensional.

A similar phenomenon takes place for $N\ge 2$, with
a tensionless brane.
Here  $A=B=C=1$, $u=\rho$ and $V_{\rm eff}(u)=0$.
The Schr\"odinger equation takes the form of Eq.~(\ref {equ})
with the above substitutions.
The total potential consists of an attractive
potential due to the induced term
and a centrifugal repulsive potential.
Because at $\rho \to 0$ the attractive potential is dominant,
one finds  properties similar to the $N=1$ case.
Heavy KK modes are suppressed on the brane ---
at distances $r\lsim  r_c$  the brane-world
gravity is four-dimensional. The only difference~ \cite {DGHS} is that
$r_c\sim \mpl/\m^2$  for $N\ge 2$.

Let us now turn to the discussion of the
case of interest when  ${\cal E}_4\ne 0$
and $A,B,C\ne 1$. Here the complete
equation (\ref {equ}) must be studied.
For the solutions that soften due to the HDO's close to  
the brane core we expect that as $\rho\to 0$, $u\sim \rho$.
Hence, to study the suppression of the wave functions
on the brane one can replace $d^2/du^2$ in (\ref {equ}) by
$d^2/d\rho^2$. The next step is to clarify the role of
the potential $V_{\rm eff}(u)$ that is nonzero when we switch on the
brane tension ${\cal E}_4\ne 0$.
Since a positive tension brane should give rise to an 
additional {\it attractive} potential in space with $N>2$, we expect 
that $V_{\rm eff}(u)$ is negative at the origin (it should tend to 
$-\infty$ at the origin if the  HDO's are not taken into account). 

The warp factors $A,B$ and $C$ contain the only 
dimensionfull parameter, $\rho_g$. So does the potential $V_{\rm eff}(u)$.
Therefore, the maximal value of the potential (if any) in the interval 
$0<\rho < \rho_g$ should be  determined by the very 
same scale, ${\rm max}\{V_{\rm eff}\}\sim  \rho_g^{-2}\,.$

If the form of the attractive potential were trustable all the way down 
to small values of the coordinate, then 
an attractive nature of the potential could  
make easier to obtain 4D  gravity on a brane as   
compared to the zero tension case. Unfortunately we cannot
draw this conclusion since the expression for the potential
is not trustable below the distance scales $\rho <\rho_*$
(see discussions in the previous section).
Although $\rho_*$ is smaller than $\rho_g$, nevertheless
this two scales can have the same order of magnitude.
Based on the discussions in the previous
section one should expect that the potential in the full theory
softens below $\rho_*$ and does not really  give rise to a substantial
attraction below that scale.
On the other hand, the potential could  give rise to some 
undesirable results. Indeed, it could produce  
a bump (a potential barrier) at some finite distance
from the core somewhere in the interval $0<\rho <\rho_g$.
For a parameter range for which this 
discussion is applicable (i.e., for $\rho_g^{-1}\ll \m$)
the hight of the bump can be of the order of  
${\rm max}\{V_{\rm eff}\}\sim  \rho_g^{-2}\,.$
A KK mode with the mass $m \gsim \rho_g^{-1}$ will not
feel the presence of of such $V_{\rm eff}$.
Its wave function will have the same properties as
in the  tensionless brane theory (i.e.  the modes with
$m > r_c^{-1}$ will be suppressed on the brane).
However, the wave-function of any KK mode with the mass  
$m \lsim \rho_g^{-1}$
will be additionally suppressed on the brane because of  the
potential barrier in $V_{\rm eff}$. The question is whether this
effect can alter the laws of 4D gravity on the brane at observable
distances. If $\rho_g$ is small this effect will certainly
spoil the emergence of 4D gravity on a brane. 
The reason is that the KK modes that are 
lighter than $\rho_g^{-1}$ will be 
additionally suppressed on the brane.
If these were  the ``active'' modes that participate
in the mediation of 4D gravity at observable distances
in the tensionless case, then having them additionally 
suppressed would change the 4D laws.    
However, if $\rho_g$ is sufficiently large the modes which are 
additionally suppressed are very light $m<\rho_g$, and, if so, 
``switching  off'' these modes won't be important for 4D gravity. 
For instance, if $\rho_g \gsim 10^{27}$ cm, 
as it happens to be the case
in the present model, gravity at observable distances
will not be noticeably different from gravity on a
tensionless brane.

Therefore, we arrive at the following qualitative
conclusion. In the worst case, gravity on the brane 
worldvolume is mediated
by the KK modes that have masses in the band
$ \rho_g^{-1} \lsim m  \lsim r_c^{-1}\,.$
Hence, at distances $r\lsim \rho_g$ the effects
of the brane tension are negligible and gravity on a
brane reproduces the known four-dimensional laws.
Moreover, in a simple case when
$\rho_g^{-1}\sim r_c^{-1}$, one can think,
qualitatively, that gravity on the brane is mediated
by a 4D graviton of  mass $$m_g \sim \rho_g^{-1} \sim r_c^{-1}\,.$$
In the present context this value is of the order of
the Hubble scale $$m_g \sim H_0 \sim 10^{-33}\,{\rm eV}\,.$$
A graviton with such a small mass is consistent with
observations~\footnote{The
discontinuity which emerges in the theory of massive gravity
at  the tree level~\cite {disc} is an artifact of the
tree-level approximation and is absent~ \cite {Arkady} in a complete
nonlinear theory. For detailed  studies of this
issue  in the context of the present model see Refs.
\cite {DDGV,Lue,Gruzinov}.}.

\section{Evading no-go arguments}

The no-go theorem by 
Weinberg \cite {Weinberg} can rule out solutions that
can be entirely understood from the low-energy point of view and
do not require fine-tuning. Here we shall briefly formulate 
these arguments and explain how
the infinite-volume theories avoid them.

A mechanism for the cancellation of the 
cosmological constant might be based on the idea 
to use scalar fields which couple to the  
vacuum energy and dynamically adjust their expectation values to 
neutralize it (a l\'a axion) (see Ref. \cite {Weinberg} for a 
review and earlier references.). 
The unbroken Lorentz invariance requires that whatever number of such 
fields are introduced,
they must transform as scalars
from the standpoint of 4D theory 
(although they may be
high-dimensional components of higher spin-fields in the underlying
UV theory).
The cancellation of the cosmological term in any such theory, with an 
arbitrarily
large but finite number of fields, requires  fine-tuning \cite {Weinberg}.
Consider a system of $n$ scalar fields $\phi_i,~\,\, i=1,2,...,n$,
 contributing
 to the vacuum energy in an arbitrary way. Then, below the scale $M_c$, 
the cosmological
equations take the conventional 4D form,
\begin{equation}
{d^2\phi_i \over dt^2}\, + \,3H\,{d\phi_i \over dt}\, + \,
\partial_{\phi_i}V(\phi)\,=\,0\,,
\label{cosmos1}
\end{equation}
and
\begin{equation}
6\,\mpl^2\, H^2 \,= \,
{{1\over 2}\,\left ({d \phi_i \over dt}\right )^2 + V(\phi)}\,,
\label{cosmos2}
\end{equation}
where $V(\phi)$ is the total potential. 

Since on the Minkowski background,
only scalars are allowed to develop  vacuum expectation values (VEV's), the
effect of all   other higher-spin fields reduces to renormalizing
parameters in $V(\phi)$ (including the $\phi$-independent constant part).
Now,  the  requirement that this system has a flat space solution, $H=0$, leaves us
with $n+1$ {\it algebraic} equations,
\begin{equation}
\partial_{\phi_i}V(\phi) =  V(\phi) = 0\,,
\label{nmore}
\end{equation}
that depend on $n$ unknowns. This system has no solution 
in general unless the parameters in $ V(\phi)$
are fine-tuned. 

To find a field configuration that
yields $H=0$ without fine-tuning
(for instance, for an arbitrary  
$\phi$-independent  part of $V(\phi)$) we have to sacrifice
at least one of the equations
\begin{equation}
\partial_{\phi_i}V(\phi) = 0,
\label{k}
\end{equation}
say at $i=k_0$. Considering the index $i$ as a ``coordinate'' in some discrete
internal space, $i=k_0$ can be regarded as a singular point in this space.
That is to say, 
we necessarily encounter a ``singular point''   trying to find a solution
with the vanishing $H$.
This  fact has a very clear-cut analogue, being considered from
the standpoint of the extra-dimensional theories. In that case $i$ can be
literally
identified with the physical extra coordinate, and the singularity mentioned 
above is just a gravitational singularity in the extra space.

Let us show this explicitly. Consider any high-dimensional
set-up that allows for such a singular solution  with the flat 4D metric
in the presence of an arbitrary bulk cosmological term, or the brane
tension.
The Einstein equation takes the form
\beq
\m^{2+N}\,\left ({\cal R}_{AB}\,-\,{1\over 2}G_{AB}\,{\cal R}\right )
= T_{AB},
\label{singular}
\eeq
where $T_{AB}$ could contain the bulk cosmological term
as well as   the brane tension. As before,
we  denote the 4D coordinates by 
$x_{\mu}$ while the extra ones by $\rho_m,~m=0,1,..,N$.

By assumption, the above equation has a solution of the form 
\beq
ds^2\,=\,A^2(\rho)\,\eta_{\mu\nu}
\,dx^\mu dx^\nu\,- \,B^2_{mn}(\rho)\, d\rho^m d\rho^n\,,
\label{flatmetric}
\eeq
which is singular at some point $\rho_m=\rho_m(k)$.
With such   metric the only components that get non-zero VEV's are the ones
that
transform as scalars under the 4D coordinate transformations ---
they have $x_{\mu}$-independent VEV's.
Performing a Fourier expansion of Eq. (\ref{flatmetric}) in terms of these
4D fields --- we shall call them 
$\phi_i$ --- and substituting them in  Eq. (\ref{singular}), we get
a set of equations similar to (\ref{nmore}). The singularity in the metric
(\ref{flatmetric}) in the language of these 4D fields is simply
a manifestation of the fact that all the equations cannot be satisfied
simultaneously. Thus, the singularity in the high-dimensional position
space
is just a ``Fourier transform'' of the fact that in the 4D language
we cannot satisfy the equation for $\phi_{k_0}$. 
This, in turn,  is a consequence
of the fact that, by requiring flat 4D space, we have over-constrained the
system. The singularity is a response to this over-constraint.
Just like in the 4D language the equations cannot be satisfied for all 
$\phi_i$'s, in the high-dimensional language the equations cannot be satisfied
at every point in the $\rho$-space.

Although one might not be able to get exactly vanishing
cosmological term without fine-tuning, one could  succeed in  
getting an acceptably small one. This is in fact what  
we  observed in the previous sections.
There is a pre-condition: this requires an {\it infinite-volume} 
extra space (or nonlocal interactions from the point of view
of 4D theory). 

To see that this is the case, return
to the 4D system described by Eqs. (\ref{cosmos1}) and (\ref{cosmos2}).
Although the system has no flat solution with $H=0$, it certainly can 
have one if we allow a non-zero curvature, $H\neq 0$, and, possibly,
time dependence of
some of the scalar fields $\phi_i$. Then all   equations
(\ref{cosmos1}) and (\ref{cosmos2}) can be satisfied simultaneously.
In the language of the  high-dimensional equation (\ref{singular}), this
means that one can smooth out the singularity, provided that 
one allows  the  four-dimensional metric to be curved. 
As it was shown in the previous sections, in 
the brane induced gravity model, this means that the 
singularity is smoothed
by letting the brane world-volume to inflate.
In order to cancel a small change
in the 4D curvature to a given accuracy $H$, by re-adjusting the VEV of a
scalar field, the latter must be as light as $H$. 
Now, as we have argued in section 2, in the
theories with finite-volume extra space, all the states that could 
re-adjust the cosmological constant have  masses of order  
$\sim 10^{-3}$ eV and cannot do the job.
On the contrary, in the 
infinite-volume models the higher-dimensional scalar components of the 
graviton KK are arbitrarily light and do
re-adjust the cosmological constant.

\section{Discussions}

In this paper we discussed a model  where 4D
gravity on the brane is obtained due to an induced 
4D Einstein-Hilbert term.
However, the arguments about the cosmological constant in general, and,
the formula (\ref {HN}) in particular, are independent
of this  mechanism.
As long as $\mpl$ does not  restrict the value
of the volume of extra space,
the arguments of the present paper will apply.

The particular model we suggest has a number of positive features:
(i) it provides a natural
explanation of the smallness of the cosmological constant; (ii)
it ensures that {\em all} gravity loops are
perfectly harmless;
(iii) finally, it preserves the logarithmic gauge coupling unification.
The model has testable predictions, for gravity 
both at sub-millimeter and Hubble
distances, and for accelerator experiments 
with the energy in the TeV range.

At this end we would like to give a brief discussion on 
the question of the cosmological evolution on the brane
(which was not the main focus of the present paper).
There are two issues to discuss: (1) How does the inflation proceed in the
early universe; (2) What would the 4D 
Friedmann-Lemaitre-Robertson-Walker (FLRW) evolution look like on the 
brane.   Let us start with (1). We expect that the inflationary 
paradigm can be incorporated in the present framework 
by using properties of the bulk. For instance, there is a possibility 
that in the early universe all the $(4+N)$-dimensions inflated 
to a very large size simultaneously before reaching the 
state that was considered in the present work. Another possibility might be 
to use the brane-inflation due to colliding branes \cite {DT}.
The later scenario provides in addition  a new mechanism  for 
the baryogenesis \cite {DGbar} 

As to the issue (2), it is certainly true that analytic solutions for 
the cosmology of matter (or radiation) dominated universe
is hard to obtain in the present context. 
However, one could gain some knowledge about these solutions using the 
arguments of the Newtonian cosmology.
In that approach all one needs is the expressions for
the potentials and forces on the brane. Since the inflation rate on the 
brane is tiny, the potentials  can be deduced 
from the corresponding flat space Green's functions. We will briefly 
summarize these discussions below.

Consider two points  located on the
brane. Their coordinates
in the worldvolume directions will be denoted by  $x$ and $x^{\prime}$,
while their $\rho$ coordinates equal to zero as they are restricted to the
brane.  We would like to discuss an  Euclidean Green
function for these two points, ${\cal G}(x-x^{\prime}, \rho=0)$.
The latter will tell us about the potential between the two sources. 
As before, it is convenient to turn to the momentum space with respect to the  
four worldvolume coordinates while
staying in the position space  with respect to $\rho$.
We will discuss  below the inverse momentum-space 
Green's function where we drop
the tensorial structure for simplicity 
\beq
{G}^{-1}(p, \rho=0)\,= \, \mpl^2
\,p^2\,+\,\m^{2+N}\,D^{-1}(p,\rho=0)\,.
\label{G}
\eeq
The first term on the right-hand side
arises due to the induced Ricci term in
Eq.~(\ref {actD}), while the second term on the  right-hand side  is
due to the bulk EH term in Eq.~(\ref {actD}). Moreover, $D(p,\rho=0)$ 
is nothing but the (Fourier transformed) Green function for the
bulk action,
\beq
D(p,\rho=0)\,=\,\int\,d^4x\, {{\rm exp}\left (ip_\mu x^\mu\right )
\over (x^2)^{2+N\over 2}}\,=\,{1\over (p^2)^{2-N\over 2}}\,
{(2\pi)^2\,\Gamma \left ({2-N\over 2}   \right )\over \Gamma \left
({2+N\over 2} \right )  }\,.
\label{Dco}
\eeq
Note that for $N\ge 2$, the momentum-space
Green's function in Eq.~(\ref {Dco}) diverges.
Thus, naively $D^{-1}(p,\rho=0) = 0$, and there is no second term on the
right-hand side of Eq.~(\ref {G}). However, as we discussed in detail 
above, this is an UV divergence
which is removed by any sensible UV regularization
introduced in the theory.
This can be done either by reintroducing a nonzero brane width~\cite {DG},
or by introducing a rigid UV cutoff in the bulk~\cite {Wagner},
or, most conveniently, by taking into account
HDO's in the bulk~\cite{Kiritsis,DGHS}. 
These  details are discussed in the
aforementioned works (see Ref.~\cite {DGHS} for a summary)
and are not important for our present purposes.
What is important here, is the fact
that $D^{-1}(p,\rho=0)\ne 0$ in the regularized theory,
and, therefore, there is an unconventional second term on the right-hand
side of Eq.~(\ref {G}). Here, we regard Eq.~(\ref {Dco})
as the {\it definition} of $D(p,\rho=0)$ in
dimensional regularization, with  ${MS}$ subtraction scheme,
where the regularization parameter $\epsilon$  is introduced as
$N\to N-2\epsilon$.
For large momenta the first term on the
right-hand side of Eq.~(\ref {G}) dominates. Hence, the UV behavior
of the inverse Green function is
\beq
{G}^{-1}(p, \rho=0)\,\simeq \, \mpl^2 \,p^2\,.
\label{UV}
\eeq
This is just  Green's function for a conventional 4D theory.
It would correspond to the
kinetic operator in 4D ``effective theory'' which is
just $\mpl^2 \,\partial_4^2$. The latter would lead to the
conventional 4D FLRW evolution on the brane.
Hence, we expect that the early cosmology in the model should be 
indistinguishable from the conventional 4D FLRW cosmology.
 
Now turn our discussion to the IR region, $p^2\ll 1/r_c$.
In this domain, the interactions on the brane become
$(4+N)$-dimensional.
This is reflected in the fact that the second term on the
right-hand side of Eq.~(\ref {G}) dominates over the $\mpl^2\,p^2$ term.
From the 4D point of view this looks as
an effect due to {\it non-local} operators in the action. 
Hence, we should expect that the present day evolution will just 
start to deviate from the conventional 4D  FLRW cosmology.

In the present paper we have suggested that a physically
acceptable solution with a small four-dimensional inflation rate can exist, 
despite the large brane tension. 
This solution represents a ``mild'' deformation of a singular 
solution with exactly flat worldvolume metric. 
Thus, in our inflating solution, as well as in the
original undeformed one, the brane tension mostly curves the nearby
region of the bulk and only slightly distorts 
four-dimensional space.
The present arguments do not exclude the existence of the 
other two solutions (discussed at the end of subsection 4.1) 
with an opposite property such that the brane tension
strongly curves the four-dimensional space. 
However, the important point is that these solutions cannot
be smoothly deformed into each other, and their realization in nature must
be determined by initial conditions in the early Universe.

\vspace{0.5cm}

{\bf Acknowledgments}
\vspace{0.1cm} \\

We would like to thank R. Emparan,
A. Vainshtein and A. Vilenkin for useful discussions. 
We thank CERN TH division and ITP at Santa Barbara 
for hospitality where parts of this work were done. 
The work of G.D. is supported in
part by a David and Lucile  Packard Foundation Fellowship
for  Science and Engineering,
by Alfred P. Sloan foundation fellowship and by NSF grant
PHY-0070787. G.G. and M.S.
are supported by  DOE grant DE-FG02-94ER408.


\begin{thebibliography}{99}

\bibitem{WeinbergPRL}
S.~Weinberg,
%``Anthropic Bound On The Cosmological Constant,''
Phys.\ Rev.\ Lett.\  {\bf 59}, 2607 (1987).
%%CITATION = PRLTA,59,2607;%%

\bibitem{cc} 
A.G. Riess et al., {  Astroph. J.} {\bf 116}, 1009 (1998);\\
S. Perlmutter et al., ``Measurements of Omega
and Lambda from 42 High-Redshift Supernovae",  astro-ph/9812133;\\
A.G. Riess, Talk Given at The Symposium ``{\it The Dark Universe:
Matter, Energy, and Gravity},'' Baltimore, April 2--5, (2001).

\bibitem{DGP} 
G.~R.~Dvali, G.~Gabadadze and M.~Porrati,
%``4D gravity on a brane in 5D Minkowski space,''
Phys.\ Lett.\ B {\bf 485}, 208 (2000)
[hep-th/0005016].
%%CITATION = HEP-TH 0005016;%%

\bibitem{DG}
G.~R.~Dvali and G.~Gabadadze,
%``Gravity on a brane in infinite-volume extra space,''
Phys.\ Rev.\ D {\bf 63}, 065007 (2001)
[hep-th/0008054].
%%CITATION = HEP-TH 0008054;%%


\bibitem{Wett}C.~Wetterich,
%``Spontaneous Compactification In Higher Dimensional Gravity,''
Phys.\ Lett.\ B {\bf 113}, 377 (1982).
%%CITATION = PHLTA,B113,377;%%


\bibitem{RS} V.~A.~Rubakov and M.~E.~Shaposhnikov,
%``Extra Space-Time Dimensions: 
%Towards A Solution To The Cosmological Constant Problem,''
Phys.\ Lett.\ B {\bf 125}, 139 (1983).
%%CITATION = PHLTA,B125,139;%%

\bibitem{Weinberg} 
S.~Weinberg,
%``The Cosmological Constant Problem,''
Rev.\ Mod.\ Phys.\  {\bf 61}, 1 (1989).
%%CITATION = RMPHA,61,1;%%

\bibitem{vilenkin} 
J.~Garriga and A.~Vilenkin,
%``Solutions to the cosmological constant problems,''
Phys.\ Rev.\ D {\bf 64}, 023517 (2001)
[arXiv:hep-th/0011262].
%%CITATION = HEP-TH 0011262;%%

\bibitem{Dines}  
K.~R.~Dienes,
%``Solving the hierarchy problem without supersymmetry 
%or extra  dimensions: An alternative approach,''
Nucl.\ Phys.\ B {\bf 611}, 146 (2001)
[arXiv:hep-ph/0104274].
%%CITATION = HEP-PH 0104274;%%

\bibitem{DV} 
G.~R.~Dvali and A.~Vilenkin,
%``Field theory models for variable cosmological constant,''
Phys.\ Rev.\ D {\bf 64}, 063509 (2001)
[arXiv:hep-th/0102142].
%%CITATION = HEP-TH 0102142;%%

\bibitem{Kiritsis}
E.~Kiritsis, N.~Tetradis and T.~N.~Tomaras,
%``Thick branes and 4D gravity,''
{  JHEP} {\bf 0108}, 012 (2001)
[hep-th/0106050].
%%CITATION = HEP-TH 0106050;%%

\bibitem{DGHS}
G.~Dvali, G.~Gabadadze, X.~Hou and E.~Sefusatti,
 ``See-saw modification of gravity,''
hep-th/0111266.
%%CITATION = HEP-TH 0111266;%%

\bibitem{ZKaku}
Z.~Kakushadze,
%``Orientiworld,''
JHEP {\bf 0110}, 031 (2001)
[arXiv:hep-th/0109054].
%%CITATION = HEP-TH 0109054;%%


\bibitem{Brown} 
S.~Corley, D.~A.~Lowe and S.~Ramgoolam,
%``Einstein-Hilbert action on the brane for the bulk graviton,''
JHEP {\bf 0107}, 030 (2001)
[arXiv:hep-th/0106067].
%%CITATION = HEP-TH 0106067;%%


\bibitem{Adler}
S. L. Adler, Phys. Rev. Lett.
{\bf 44} 1567 (1980);
%A FORMULA FOR THE INDUCED GRAVITATIONAL
%CONSTANT.
Phys. Lett. {\bf B95}, 241 (1980);
%EINSTEIN
%GRAVITY AS A SYMMETRY BREAKING EFFECT IN QUANTUM FIELD THEORY
Rev. Mod. Phys. {\bf 54} 729
(1982); Erratum-ibid. {\bf 55}, 837 (1983).

\bibitem{Zee} A. Zee,
%CALCULATING NEWTON'S GRAVITATIONAL CONSTANT IN INFRARED
%STABLE YANG-MILLS THEORIES.
Phys. Rev. Lett. {\bf 48} 295 (1982).



%\bibitem{Rubakov}V.~A.~Rubakov and M.~E.~Shaposhnikov,
%``Do We Live Inside A Domain Wall?,''
%Phys.\ Lett.\ B {\bf 125},  136 (1983).
%%CITATION = PHLTA,B125,136;%%


\bibitem{mishagia}
G.~R.~Dvali and M.~A.~Shifman,
%``Dynamical compactification as a mechanism 
%of spontaneous supersymmetry  breaking,''
Nucl.\ Phys.\ B {\bf 504}, 127 (1997)
[hep-th/9611213].
%%CITATION = HEP-TH 9611213;%%


\bibitem{DGPsusy} 
G.~R.~Dvali, G.~Gabadadze and M.~Porrati,
%``Metastable gravitons and infinite volume extra dimensions,''
Phys.\ Lett.\ B {\bf 484}, 112 (2000)
[hep-th/0002190];
%%CITATION = HEP-TH 0002190;%%
%G.~R.~Dvali, G.~Gabadadze and M.~Porrati,
%``A comment on brane bending and ghosts in theories
%with infinite extra  dimensions,''
Phys.\ Lett.\ B {\bf 484}, 129 (2000)
[hep-th/0003054].
%%CITATION = HEP-TH 0003054;%%

\bibitem{Wittensusy} 
E.~Witten,
``The cosmological constant from the viewpoint of string theory,''
hep-ph/0002297.
%%CITATION = HEP-PH 0002297;%%



\bibitem{DGKN1}
G.~R.~Dvali, G.~Gabadadze, M.~Kolanovic and F.~Nitti,
%``Scales of gravity,''
hep-th/0106058;\\
%%CITATION = HEP-TH 0106058;%%
Phys.\ Rev.\ D {\bf 64}, 084004 (2001)
[hep-ph/0102216].
%%CITATION = HEP-PH 0102216;%%

\bibitem{Sundrum}
R.~Sundrum,
%``Towards an effective particle-string
%resolution of the cosmological  constant problem,''
{JHEP} {\bf 9907}, 001 (1999)
[hep-ph/9708329].
%%CITATION = HEP-PH 9708329;%%



\bibitem{Moffat}
J.~W.~Moffat,
%``Quantum field theory solution to the gauge 
%hierarchy and cosmological  constant problems,''
arXiv:hep-ph/0003171.
%%CITATION = HEP-PH 0003171;%%


\bibitem{Wagner}
M.~Carena, A.~Delgado, J.~Lykken, S.~Pokorski, M.~Quiros and C.~E.~Wagner,
%``Brane effects on extra dimensional scenarios: A tale of two
%gravitons,''
Nucl.\ Phys.\ B {\bf 609}, 499 (2001)
[hep-ph/0102172].
%%CITATION = HEP-PH 0102172;%%

\bibitem{Gruzinov}
A.~Gruzinov,
 ``On the graviton mass,''
astro-ph/0112246.
%%CITATION = ASTRO-PH 0112246;%%

\bibitem{Horowitz}
G.~T.~Horowitz and A.~Strominger,
%``Black Strings And P-Branes,''
Nucl.\ Phys.\ B {\bf 360}, 197 (1991).
%%CITATION = NUPHA,B360,197;%%


\bibitem{Gregory}
R.~Gregory,
%``Cosmic p-Branes,''
Nucl.\ Phys.\ B {\bf 467}, 159 (1996)
[hep-th/9510202].
%%CITATION = HEP-TH 9510202;%%

\bibitem{Emparan}
C.~Charmousis, R.~Emparan and R.~Gregory,
%``Self-gravity of brane worlds: A new hierarchy twist,''
JHEP {\bf 0105}, 026 (2001)
[hep-th/0101198].
%%CITATION = HEP-TH 0101198;%%



\bibitem{Zhou}
B.~Zhou and C.~J.~Zhu,
%``A study of brane solutions in D-dimensional coupled gravity system,''
Commun.\ Theor.\ Phys.\  {\bf 32}, 507 (1999)
[arXiv:hep-th/9903118];\\
%%CITATION = HEP-TH 9903118;%%
B.~Zhou and C.~J.~Zhu,
%``The complete black brane solutions in 
%D-dimensional coupled gravity  system,''
arXiv:hep-th/9905146.
%%CITATION = HEP-TH 9905146;%%



\bibitem{Vilenkin1}
A.~Vilenkin,
%``Gravitational Field Of Vacuum Domain Walls And Strings,''
Phys.\ Rev.\ D {\bf 23} (1981) 852.
%%CITATION = PHRVA,D23,852;%%


\bibitem{Sikivie}
A.~Vilenkin,
%``Gravitational Field Of Vacuum Domain Walls,''
Phys.\ Lett.\  {\bf B133}, 177 (1983);\\
J.~Ipser and P.~Sikivie,
%``The Gravitationally Repulsive Domain Wall,''
Phys.\ Rev.\  {\bf D30}, 712 (1984).

\bibitem{CK1} 
A.~G.~Cohen and D.~B.~Kaplan,
%``The Exact Metric About Global Cosmic Strings,''
Phys.\ Lett.\ B {\bf 215}, 67 (1988).
%%CITATION = PHLTA,B215,67;%%

\bibitem{GregoryCK1}
R.~Gregory,
%``Non-singular global strings,''
Phys.\ Rev.\ D {\bf 54}, 4955 (1996)
[gr-qc/9606002].
%%CITATION = GR-QC 9606002;%%

\bibitem{Cho}
I.~Cho,
%``Inflation and nonsingular spacetimes of cosmic strings,''
Phys.\ Rev.\ D {\bf 58}, 103509 (1998)
[gr-qc/9804086].
%%CITATION = GR-QC 9804086;%%

\bibitem{CK2} 
A.~G.~Cohen and D.~B.~Kaplan,
%``Solving the hierarchy problem with noncompact extra dimensions,''
Phys.\ Lett.\ B {\bf 470}, 52 (1999)
[hep-th/9910132].
%%CITATION = HEP-TH 9910132;%%

\bibitem{Erich} 
A.~Chodos, E.~Poppitz and D.~Tsimpis,
%``Nonsingular deformations of singular
%compactifications, the  cosmological constant, and the hierarchy
%problem,''
Class.\ Quant.\ Grav.\  {\bf 17}, 3865 (2000)
[hep-th/0006093].
%%CITATION = HEP-TH 0006093;%%

\bibitem{VC}R.~Cordero and A.~Vilenkin,
%``Stealth branes,''
arXiv:hep-th/0107175.
%%CITATION = HEP-TH 0107175;%%

\bibitem{Dick} R.~Dick,
%``Brane worlds,''
Class.\ Quant.\ Grav.\  {\bf 18}, R1 (2001)
[arXiv:hep-th/0105320];\\
%%CITATION = HEP-TH 0105320;%%
R.~Dick,
%``Standard cosmology in the DGP brane model,''
Acta Phys.\ Polon.\ B {\bf 32}, 3669 (2001)
[arXiv:hep-th/0110162].
%%CITATION = HEP-TH 0110162;%%

\bibitem{Cedric}
C.~Deffayet,
%``Cosmology on a brane in Minkowski bulk,''
Phys.\ Lett.\ B {\bf 502}, 199 (2001)
[arXiv:hep-th/0010186].
%%CITATION = HEP-TH 0010186;%%


\bibitem{Niles}
S.~Forste, Z.~Lalak, S.~Lavignac and H.~P.~Nilles,
%``A comment on self-tuning and vanishing 
%cosmological constant in the  brane world,''
Phys.\ Lett.\ B {\bf 481}, 360 (2000)
[arXiv:hep-th/0002164].
%%CITATION = HEP-TH 0002164;%%

\bibitem{LindeT} 
A.~D.~Linde,
%``Monopoles as big as a universe,''
Phys.\ Lett.\ B {\bf 327}, 208 (1994)
[astro-ph/9402031].
%%CITATION = ASTRO-PH 9402031;%%

\bibitem{VilenkinT}
A.~Vilenkin,
%``Topological inflation,''
Phys.\ Rev.\ Lett.\  {\bf 72}, 3137 (1994)
[hep-th/9402085].
%%CITATION = HEP-TH 9402085;%%


\bibitem{Borut}
B.~Bajc and G.~Gabadadze,
%``Localization of matter and cosmological constant
%on a brane in anti de  Sitter space,''
Phys.\ Lett.\ B {\bf 474}, 282 (2000)
[hep-th/9912232].
%%CITATION = HEP-TH 9912232;%%

\bibitem{Csaki} 
C.~Csaki, J.~Erlich, T.~J.~Hollowood and Y.~Shirman,
%``Universal aspects of gravity localized on thick branes,''
Nucl.\ Phys.\ B {\bf 581}, 309 (2000)
[hep-th/0001033].
%%CITATION = HEP-TH 0001033;%%

\bibitem{disc}
Y. Iwasaki, Phys. Rev. {\bf D2}, 2255 (1970);\\
H. van Dam and M. Veltman, Nucl. Phys.
{\bf B22}, 397 (1970);\\
V.~I.~Zakharov, JETP Lett. {\bf 12}, 312 (1970).

\bibitem{Arkady} 
A.~I.~Vainshtein, Phys. Lett.
{\bf 39B}, 393 (1972).

\bibitem{DDGV}
C.~Deffayet, G.~R.~Dvali, G.~Gabadadze and A.~I.~Vainshtein,
%``Nonperturbative continuity in graviton 
%mass versus perturbative  discontinuity,''
Phys.\ Rev.\ D {\bf 65}, 044026 (2002)
[arXiv:hep-th/0106001].
%%CITATION = HEP-TH 0106001;%%



\bibitem{Lue} A.~Lue,
%``Cosmic strings in a brane world theory with metastable gravitons,''
arXiv:hep-th/0111168.
%%CITATION = HEP-TH 0111168;%%

%\bibitem{Duff} 
%M.~J.~Duff, R.~R.~Khuri and J.~X.~Lu,
%``String solitons,''
%Phys.\ Rept.\  {\bf 259}, 213 (1995)
%[hep-th/9412184].
%%CITATION = HEP-TH 9412184;%%

%\bibitem{Oz} 
%O.~Aharony, S.~S.~Gubser, J.~Maldacena, H.~Ooguri and Y.~Oz,
%``Large N field theories, string theory and gravity,''
%Phys.\ Rept.\  {\bf 323}, 183 (2000)
%[hep-th/9905111].
%%CITATION = HEP-TH 9905111;%%



\bibitem{zura2}
O.~Corradini, A.~Iglesias, Z.~Kakushadze and P.~Langfelder,
%``Gravity on a 3-brane in 6D bulk,''
Phys.\ Lett.\ B {\bf 521}, 96 (2001)
[hep-th/0108055].
%%CITATION = HEP-TH 0108055;%%


\bibitem{DT}
G.~R.~Dvali and S.~H.~Tye,
%``Brane inflation,''
Phys.\ Lett.\ B {\bf 450}, 72 (1999)
[arXiv:hep-ph/9812483].
%%CITATION = HEP-PH 9812483;%%


\bibitem{DGbar}
G.~R.~Dvali and G.~Gabadadze,
%``Non-conservation of global charges in 
%the brane universe and  baryogenesis,''
Phys.\ Lett.\ B {\bf 460}, 47 (1999)
[arXiv:hep-ph/9904221].
%%CITATION = HEP-PH 9904221;%%


\end{thebibliography}
\end{document}